\documentclass{birkau}

\usepackage{amsmath,amssymb,url}
\usepackage{amssymb}
\usepackage [all]{xy}
\usepackage{mathrsfs}
\usepackage{mathtools}
\usepackage{stmaryrd}
\usepackage{lmodern}
\usepackage{amssymb}
\usepackage{graphicx}
\usepackage [all]{xy}
\usepackage{mathrsfs}
\usepackage{lmodern}

\usepackage{tikz-dependency}

\makeatletter
\def\new@fontshape{}
\makeatother
\usepackage{gb4e}
\noautomath

\numberwithin{equation}{section}

\theoremstyle{plain}
\newtheorem{theorem}{Theorem}[section]
\newtheorem{lemma}[theorem]{Lemma}
\newtheorem{proposition}[theorem]{Proposition}

\theoremstyle{definition}
\newtheorem{definition}[theorem]{Definition}

\newtheorem{remark}[theorem]{Remark}
\newtheorem{example}[theorem]{Example}

\DeclareMathOperator{\depth}{depth}
\DeclareMathOperator{\dom}{dom}
\DeclareMathOperator{\Root}{root}

\newcommand{\Loc}{\mathbf{Loc}}
\newcommand{\Tree}{\mathbf{Tree}}

\newcommand{\Lcopy}{L_{\mathrm{copy}}}
\newcommand{\Lsqua}{L_{\mathrm{squa}}}
\newcommand{\Lresp}{L_{\mathrm{resp}}}
\newcommand{\Lmult}{L_{\mathrm{mult}}}
\newcommand{\LDyck}{L_{\mathrm{Dyck}}}
\newcommand{\Lncopy}{L_{n\operatorname{-copy}}}
\newcommand{\Ltcopy}{L_{2\operatorname{-copy}}}
\newcommand{\Lmix}{L_{\mathrm{mix}}}

\newcommand{\LC}{\mathsf{LC}}
\newcommand{\PR}{\mathsf{PR}}
\newcommand{\CF}{\mathsf{CF}}
\newcommand{\UN}{\mathsf{UN}}
\newcommand{\FN}{\mathsf{FN}}
\newcommand{\RG}{\mathsf{RG}}
\newcommand{\LS}{\mathsf{LS}}

\newcommand{\Sb}{\mathit{Sb}}
\newcommand{\Ob}{\mathit{Ob}}
\newcommand{\Dt}{\mathit{Dt}}
\newcommand{\Ad}{\mathit{Ad}}
\newcommand{\Md}{\mathit{Md}}

\begin{document}


\title[Anti-Context-Free Languages]{Anti-Context-Free Languages}
\author[C. Card\'o]{Carles Card\'o}
\address{Departament de Ci\`encies de la Computaci\'o, \\Campus Nord, Edifici Omega, Jordi Girona Salgado 1-3. 08034  \\Universitat Polit\`ecnica de Catalunya \\ Barcelona, \\Catalonia (Spain)}
\urladdr{http://www.cs.upc.edu}
\email{cardocarles@gmail.com}

\thanks{To appear at Journal of Automata, Languages and Combinatorics.}

\begin{abstract} 
Context-free languages can be characterized in several ways. This article studies projective linearisations of languages of simple dependency trees, i.e., dependency trees
in which a node can govern at most one node with a given syntactic function.
We prove that the projective linearisations of local languages of simple dependency trees coincide with the context-free languages.

Simple dependency trees suggest alternative dual notions of locality and projectivity, which permits defining a dual language for each context-free language. We call this new class of languages anti-context-free. These languages are related to some linguistic constructions exhibiting the so-called cross-serial dependencies that were historically important for the development of computational linguistics. We propose that this duality could be a relevant linguistic phenomenon.
\end{abstract}

\maketitle


\section{Introduction} \label{Introduction}

\subsection{Historical and linguistic motivation}

From the beginning of the history of computational linguistics, there was a suspicion that, although context-free grammars could describe a large part of natural languages, they were insufficient \cite{chomsky2002syntactic}.
After some discussions over the next decades (see \cite{pullum1982natural} for state of the art at that time), it was confirmed that some natural languages present non-context-free constructions \cite{shieber1987evidence}.
In this respect, let us consider a pair of formal languages for a fixed alphabet $\Sigma$:
\begin{align}
\Lsqua &=\{ a_1^2 \cdots a_n^2 \,|\, a_1, \ldots, a_n \in \Sigma, n\geq 0\},\\
\Lcopy &=\{ xx \,|\, x\in \Sigma^*\},
\end{align}
which here we will call the \emph{language of squares} and the \emph{copy language}, respectively. The first is a context-free, indeed regular, language. It is a mathematical abstraction of a chain of subordinate clauses in English. For example:
\begin{exe}
\ex \label{English}
\ldots \ that John$^a$ saw$^a$ Peter$^b$ help$^b$ Mary$^c$ read$^c$.
\end{exe}
The second is not context-free and represents a chain of subordinate clauses in Dutch \cite{bresnan1987cross}. For example:
\begin{exe}
\ex \label{Dutch}
\gll
\ldots \ dat Jan$^a$ Piet$^b$ Marie$^c$ zag$^a$ helpen$^b$ lezen$^c$.
\\ \ldots \ that Jan Piet Marie saw help read
\\
\trans
`\ldots \ that J. saw P. help M. read.'
\end{exe}
This example is said to exhibit \emph{cross-serial dependencies}.
The same crossing configuration is found in Swiss-German. The links are visible since Swiss-German has case marking:
\begin{exe}
\ex
\gll
Jan s\"ait das mer d'chind em Hans es huus l\"ond h\"alfe aastriiche.
\\
John said that the children$_\mathit{ACC}$ $\,$ Hans$_\mathit{DAT}$ the house$_{ACC}$ let help paint.
\\
\end{exe}
In the sentences (\ref{English}) and (\ref{Dutch}), the abstraction into the formal language follows from the argument structure; that is, we mark with the same letter the subject and the related verb.
However, the case of Swiss-German can be formally reduced into the copy language by intersecting the whole Swiss-German language with a regular language and reducing by a homomorphism of monoids according to the case mark \cite{kallmeyer2010parsing}. Since context-free languages are closed under intersections with regular languages and homomorphisms \cite{hopcroft1979introduction}, Swiss-German is not context-free.

Another pair of formal languages that we want to consider are the \emph{multiple $abc$ language} and the \emph{respectively $abc$ language}:
\begin{align}
L_{\mathrm{mult}} &=\{ (abc)^n \,|\, n\geq 0 \},\\
L_{\mathrm{resp}} &=\{ a^n b^n c^n \,|\, n\geq 0\}.
\end{align}
The first is a context-free, indeed regular language, but not the second. The first corresponds to simple coordination, as in sentence~(\ref{frasescoordinacio})a, while the second is a mathematical idealization of the \emph{respectively} construction, as in sentence~(\ref{frasescoordinacio})b:
\begin{exe} \ex \label{frasescoordinacio}
\begin{xlista}
\ex
Jean$^a$ seems German$^b$ but he is French$^c$, Pietro$^a$ seems Russian$^b$ but he is Italian$^c$ and Peter$^a$ seems Belgian$^b$, but he is English$^c$.
\ex
Jean$^a$, Pietro$^a$ and Peter$^a$ seem German$^b$, Russian$^b$ and Belgian$^b$, respectively, but they are French$^c$, Italian$^c$ and English$^c$.
\end{xlista}
\end{exe}
Respectively constructions, which also exhibit cross-serial dependencies, had already been considered as initial counterexamples against context-freeness of English with sentences as:\footnote{See \cite{bar1960finite} for (\ref{frasescoordinacio2})a, and \cite{kac1987simultaneous} for (\ref{frasescoordinacio2})b.}
\begin{exe} \ex \label{frasescoordinacio2}
\begin{xlista}
\ex
John, Mary and David are a widower, a widow and a widower respectively.
\ex
This land and these woods can be expected to rend itself and sell themselves respectively.
\end{xlista}
\end{exe}
However, some linguists considered such constructions too artificial, the examples in Swiss-German being more transparent.

In any case, given linguistic data, formal grammars should be able to generate such cross-serial dependencies. The strategy was to give just a little extra power to context-free grammars to incorporate the copy language, and similar constructions, which was called \emph{mildly context-sensitivity} \cite{joshi1985tree}. A little zoo of formal systems was proposed during the following years: \emph{tree adjoining grammars TAG} \cite{joshi1985tree}; \emph{linear indexed grammars LIG} \cite{gazdar1988applicability}; $q$-\emph{linear context-free rewriting systems $q$-LCFRS} \cite{vijay1987characterizing}; $q$-\emph{multiple context-free grammars $q$-MCFG} \cite{seki1991multiple}. Some of these formalisms are equivalent; for example, linear indexed grammars are equivalent to tree adjoining grammars, and $q$-LCFRS are equivalent to $q$-MCFG, for any $q$. Moreover, some formalisms are subsumed by others; for example, linear context-free rewriting systems are more expressive than linear indexed grammars. See \cite{kallmeyer2010parsing} for a survey on the issue. Some linguists assume that natural language's weak capacity could be very close to MCFG, although the debate continues. On the one hand, Kanazawa and Salvati \cite{kanazawa2012mix} invite to refine this claim, since the \emph{mix language}, $\Lmix=$ $\{ x \in \{a,b,c\}^* $ $\mid |x|_a=|x|_b=|x|_c\}$, is generable by a $2$-MCFG, but such language is thought not pertinent for human language \cite{bach1988categorial, bach1981discontinous}.
On the other hand, we have to consider the possibility of non-semi-linear constructions, such as in Yoruba or Old-Georgian, which overpasses the generative capacity of MCFG, see \cite{pmlr-v21-clark12a}.

Historic considerations aside, let us examine the four languages above $\Lcopy$, $\Lsqua$, $\Lmult$, and $\Lresp$, and redefine them with the following notation. For a string $x=x_1\cdots x_n \in \Sigma^*$, where $\Sigma$ is any alphabet and $x_1, \ldots, x_n \in \Sigma$, we write $x^{\uparrow m}=x^m$ (that is, the usual power of strings) and $x^{\downarrow m}=x_1^m \cdots x_n^m$.
Consider the following table:
\begin{center}
\begin{tabular}{ c c }
\bf Context-Free & \bf Non-Context-Free\\
$\Lsqua=\{ x^{\downarrow 2} \mid x \in \Sigma^*\}$ & $\Lcopy=\{ x^{\uparrow 2} \mid x \in \Sigma^*\}$ \\
$\Lmult=\{ (abc)^{\uparrow n} \mid n\geq 0\}$ & $\Lresp=\{ (abc)^{\downarrow n} \mid n\geq 0\}$
\end{tabular}
\end{center}
There is a duality between context-free and non-context-free languages under the operations $\uparrow$ and $\downarrow$.
One of the goals of this article is to formalize that duality giving a theoretical framework. Beyond these simple examples, we will see that there exists a dual language for each context-free language. Thus, it is natural to consider a dual class of context-free languages. We call this class \emph{anti-context-free}. Such languages were introduced in \cite{cardo2018algebraic} in the wider framework of \emph{Algebraic Dependency Grammar}, while the notion of duality as a linguistic phenomenon was already suggested in \cite{cardo2016algebraic}. However, and importantly, here we recast definitions, theorems, and proofs without the heavy algebraic apparatus given earlier in \cite{cardo2018algebraic}.

\subsection{The dependency approach} \label{TheDependencyApproach}

The phenomenon of duality has only been described in the framework of \emph{dependency grammar}.
Dependency structures are an alternative form of syntactic analysis introduced by the French grammarian L. Tesni{\`e}re \cite{tesniere1959elements}. Unlike constituent analysis, a dependency structure is rather a relational structure over the words of a sentence. For example, in the sentence \emph{the boy caught frogs}, the word \emph{caught} governs the word \emph{frogs} as object ($\Ob$), that forms the dependence:
 \begin{center}
\begin{dependency}[arc edge, arc angle=60, text only label, label style={above}]
\begin{deptext}[column sep=.6cm]
caught \& frogs \\
\end{deptext}
\depedge{1}{2}{\it Ob}
\end{dependency}
\end{center}
Joining together a set of dependencies we have a dependency structure:
 \begin{center}
\begin{dependency}[arc edge, arc angle=60, text only label, label style={above}]
\begin{deptext}[column sep=.6cm]
the \& boy \& caught \& frogs \\
\end{deptext}
\depedge{3}{2}{\it Sb}
\depedge{3}{4}{\it Ob}
\depedge{2}{1}{\it Dt}
\end{dependency}
\end{center}
In what follows we will use the abbreviations $\Sb$ subject, $\Dt$ determiner, $\Ob$ object, $\Ad$ adjective, $\Md$ modifier, $\mathit{Cj}$ conjunction, $\mathit{Co}$ coordination. Other authors use different nomenclature for such syntactic functions, but this is not relevant for the examples in this article.

Dependency structures can adopt several topologies, such as networks for semantics or directed acyclic graphs for some syntactic representations. However, the most fundamental is the tree-shape \cite{nivre2005dependency}.
There is not a unique style of parse trees. Several treebanks allow a node to govern two or more children with the same syntactic function. This configuration arises in coordination constructions, a particularly controversial issue in dependency grammar. A possible style of analysis for coordination is:
 \begin{center}
\begin{dependency}[arc edge, arc angle=60, text only label, label style={above}]
\begin{deptext}[column sep=.4cm]
John \& caught \& \,\, frogs \& dragonflies \& and \& butterflies\\
\end{deptext}
\depedge{2}{1}{\it Sb}
\depedge{2}{3}{\it \qquad \,\,\, Ob}
\depedge{2}{4}{\it \,\,\,\, Ob}
\depedge{2}{6}{\it Ob}
\depedge{2}{5}{\it \,\,\,\,\, Cj}
\end{dependency}
\end{center}
Another variant is:
 \begin{center}
\begin{dependency}[arc edge, arc angle=60, text only label, label style={above}]
\begin{deptext}[column sep=.4cm]
John \& caught \& \,\, frogs \& dragonflies \& and \& butterflies\\
\end{deptext}
\depedge{2}{1}{\it Sb}
\depedge{5}{3}{\it Co \qquad}
\depedge{5}{4}{\it Co \qquad \quad}
\depedge{5}{6}{\it Co}
\depedge{2}{5}{\it Ob}
\end{dependency}
\end{center}
In the Russian linguistic tradition \cite{melcuk1988dependency}, a unique coordinator function $\mathit{Co}$ is needed to chain the coordinants. For example:
 \begin{center}
\begin{dependency}[arc edge, arc angle=60, text only label, label style={above}]
\begin{deptext}[column sep=.4cm]
John \& caught \& frogs \& dragonflies \& and \& butterflies\\
\end{deptext}
\depedge{2}{1}{\it Sb}
\depedge{2}{3}{\it Ob}
\depedge{3}{4}{\it Co}
\depedge{4}{6}{\it Co}
\depedge{6}{5}{\it Cj \,\,\,}
\end{dependency}
\end{center}
These are only some variants. See \cite{popel2013coordination} for a comparison among styles of parsing from different treebanks. 

In this article, we will use \emph{simple dependency trees}, defined as a partial mapping from syntactic addresses to the vocabulary, rather than graphs, see Definition~\ref{DefinitionTree}, Section~\ref{Definitions}. This format is algebraically compact but only allows at most one child governed by a function. In particular, simple dependency trees only model a part of dependency trees. For coordination constructions, the reader with linguistic interests should adopt, at least for this article, the last commented style.
However, leaving aside the linguistic motivation of the current and the last section, this article focuses on the formal aspects of duality and formal language theory. In-depth linguistic discussions are beyond the scope of this article.

\subsection{Projective word-orderings} \label{ProjectiveIntro}

\begin{figure}[tb] 
\centering
\includegraphics[height=58mm]{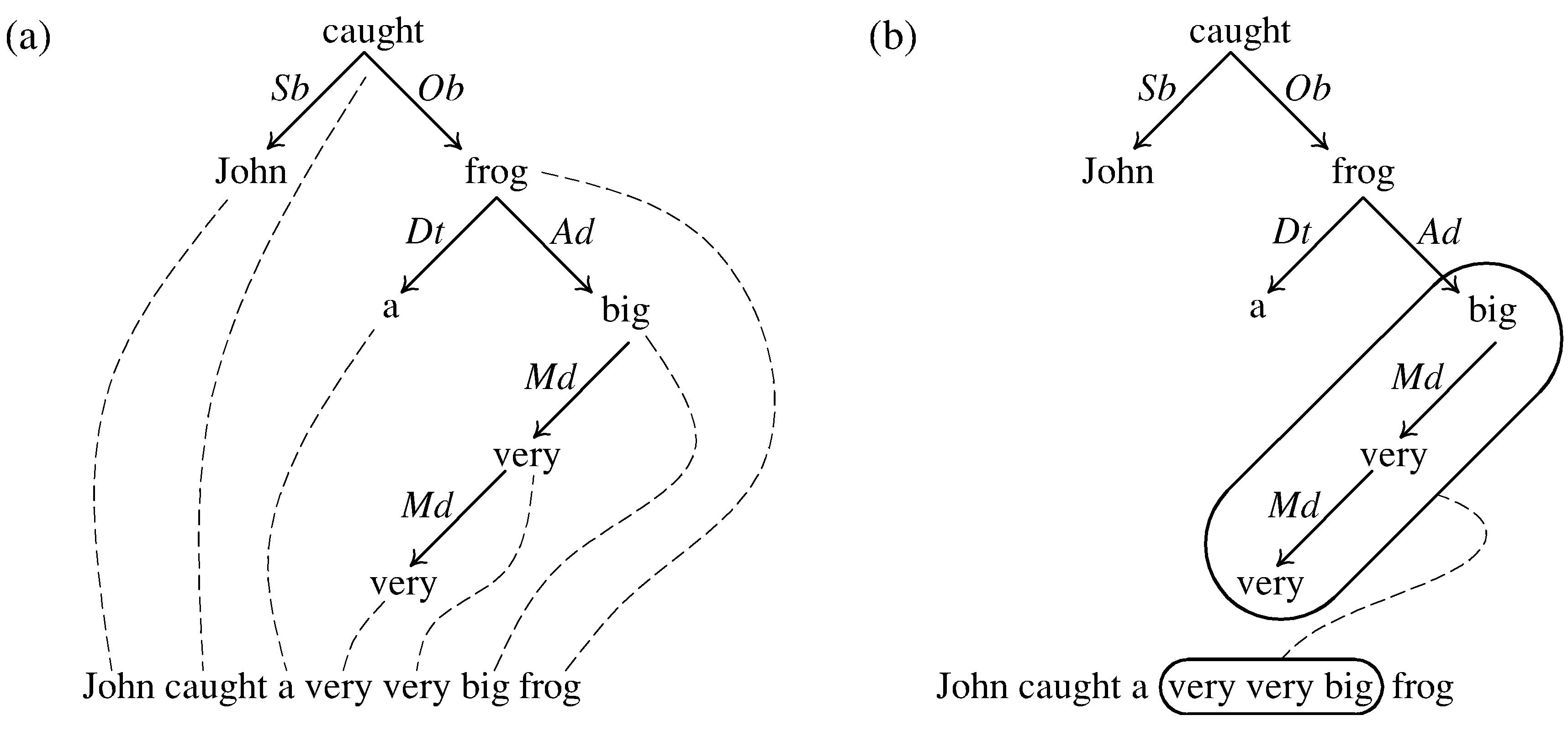}
\caption{(a) Dependency tree and a projective word-ordering.  (b) An example of subtree transformed into a substring.}\label{Fig1}
\end{figure}

The above dependency trees show the tree and the phrase simultaneously. However, we can separate the structures. A word-ordering is a relationship between the dependency tree and the places in the linear order of the sentence. An important type of word-ordering is the \emph{projective word-ordering}. Projective word-orderings can be represented graphically by lines from the dependency tree to the linear order without intersecting (this needs additional constraints that we do not address here). See Figure~\ref{Fig1}(a).
There exist several equivalent characterizations of projectivity. Here we adapt the following from \cite{kuhlmann2010dependency}:
\emph{A word-ordering is projective if it transforms (total) subtrees of the dependency tree into substrings}. See an example in Figure~\ref{Fig1}(b).

Statistics from treebanks show that most word-orderings are projective. There are, of course, remarkable exceptions. The Dutch sentence (\ref{Dutch}) is an example:
 \begin{center}
\begin{dependency}[arc edge, arc angle=70, text only label, label style={above}]
\begin{deptext}[column sep=.4cm]
\ldots \ dat \& Jan \& Piet \&  Marie \& zag \& helpen \&  lezen\\
\end{deptext}
\depedge{5}{6}{\it Ob\quad\quad\quad}
\depedge{6}{7}{\it Ob \quad\quad\quad}
\depedge{5}{2}{\it Sb}
\depedge{6}{3}{\it Sb}
\depedge{7}{4}{\it Sb}
\end{dependency}
\end{center}
A major issue in dependency grammar is how to formalize these non-projective word-orderings. 

\subsection{Structure of the article}

Once we have fixed notation and preliminary definitions in Section~\ref{Definitions}, we introduce in Section~\ref{Locality} the class of \emph{local simple dependency tree languages}, and, in Section~\ref{Projectivity}, the class of \emph{recursive projective linearisations}. Next, we prove that the class of context-free languages is equivalent to the local simple dependency tree languages projectively linearised.
There exists a classical result in this direction for rewriting systems: It is proved that the set of derivation trees of a context-free grammar is local \cite{comon2007tree}. In terms of the so-called dependency systems, a similar result was already proved \cite{gross1964equivalence, hays1964dependency, gaifman1965dependency, robinson1970dependency}.
However, here we offer a version for tree languages and projective linearisations as independent objects.

The definition of simple dependency trees invites us to consider dual alternatives. So we can define for each tree language a dual tree language, and the same occurs in the case of linearisations, Section~\ref{AntiClasses}.
We will prove that by taking the dual class of local languages and the dual class of projections, we obtain context-free languages again. However, by taking the dual form of only one of these classes, we obtain the new class so-called \emph{anti-context-free languages}.
Section~\ref{AntiContextFree} defines, shows examples, and studies the main properties of those languages. We have, for instance, that the respectively language and the copy language are anti-context-free.
Finally, we resume the linguistic discussion in Section~\ref{Discussion}.

\section{Notation and Definitions} \label{Definitions}

Given a finite set $A$, $A^*$ denotes the free monoid over $A$, the elements of which are called \emph{strings}. The \empty{empty string} is denoted by $\varepsilon$. The \emph{length} of a string $x$ is denoted by $|x|$, and $|x|_a$ stands for the number of occurrences of a letter $a$ in the string $x$. When necessary, we will use the multiplicative notation, $x\cdot y=xy$.
Given the string $x=x_1\cdots x_n \in A^*$, with $x_1, \ldots, x_n \in A$, the reversed string is defined as $x^R=x_n \cdots x_1$. A string $y$ is said to be a \emph{substring} of $y'$ if there are strings $y,z$ such that $xyz=y'$. Given a permutation $\rho: \{1, \ldots, m\} \longrightarrow \{1, \ldots, m\}$, and given $m$ strings $x_1, \ldots, x_m$, we will write $\rho(x_1, x_2, \ldots, x_n)$ for the string $x_{\rho(1)}x_{\rho(2)}\cdots x_{\rho(m)}$.
Since we will use simultaneously two free monoids $\zeta^*$ and $\Sigma^*$ for different uses, we reserve $\varepsilon$ for the empty string of $\Sigma^*$, and $1$ for the empty string of $\zeta^*$.
The domain of a partial mapping $f$ is denoted by $\dom f$.

\begin{definition} \label{DefinitionTree} Let $\Sigma$ and $\zeta$ be finite sets, called \emph{vocabulary} and \emph{syntactic functions}, respectively.
A \emph{simple dependency tree} is a partial mapping $S: \zeta^* \longrightarrow \Sigma$ with finite domain.
The set of simple dependency trees over $\zeta$ and $\Sigma$ is denoted by $\Tree_{\zeta, \Sigma}$. When the context permits, we will write $\Tree$, for short.
The depth of a simple dependency tree is defined as $\depth(S)=\max \{ |x| \mid x \in \dom S\}$.
Simple dependency trees with depth zero, are called \emph{atomic}, written $a^\bullet$, where $a\in \Sigma$, that is, $\dom S=\{1\}$ and $a^\bullet(1)=a$. We accept the empty set as a simple dependency tree with $\depth(\emptyset)=0$.
In what follows, we will say \emph{tree} to refer to a simple dependency tree according to this definition when no confusion can arise.
\end{definition}

\begin{remark}
If we take the prefix closure of the domain of $S \in \Tree$, then $S$ can be interpreted as a tree. Every vertex can be identified by a unique \emph{Gorn address} $x \in \zeta^*$. When $x \in \dom S$, we label the vertex as $S(x)$. Prefixes not in the domain are viewed as unlabelled vertices.
Definition~\ref{DefinitionTree} implies that given a syntactic function $\alpha \in \zeta$, and given a vertex $x$ there exists at most one child $y$ such that $x$ governs $y$ by $\alpha$. So, the class of simple dependency trees only captures a certain kind of dependency tree.
\end{remark}

\begin{example} \label{ExampleTree}
Consider the set of syntactic functions $\zeta=\{ \Sb, \Ob, \Dt, \Ad, \Md \}$, recall the nomenclature given in Section~\ref{TheDependencyApproach}. Let the vocabulary be $\Sigma=\{$the, boy, caught, frogs$\}$. The tree: 
\begin{align*}
S(x)=\begin{cases} \mbox{caught} & \mbox{ if } x=1,\\
\mbox{John} & \mbox{ if } x=\Sb,\\
\mbox{frog} & \mbox{ if } x=\Ob,\\
\mbox{a} & \mbox{ if } x=\Ob \cdot \Dt,\\
\mbox{big} & \mbox{ if } x=\Ob \cdot \Ad,\\
\mbox{very} & \mbox{ if } x=\Ob \cdot \Ad \cdot \Md,\\
\mbox{very} & \mbox{ if } x=\Ob \cdot \Ad \cdot \Md \cdot \Md.
\end{cases}
\end{align*}
can be represented as in Figure~\ref{Fig1}(a). 
Gorn addresses must be read in reverse order to make linguistic sense. For example, ``the determiner of the object'' is the path $\Ob \cdot \Dt$.
\end{example}

\begin{definition} \label{Operators} Given a tree $S$, we introduce the operators  
$[S]^p,\,\, \overline{S}, \,\, S|\varphi,\,\, \varphi |S\,\,: \Tree_{\zeta,\Sigma} \longrightarrow \Tree_{\zeta,\Sigma}$,
where $p\geq 0$ is an integer and $\varphi  \in \zeta^*$, defined as follows:
\begin{itemize}
\item[(i)] $[S]^p(x)=\begin{cases}S(x) & \mbox{ if } |x|\leq p, \\ \mbox{not defined } & \mbox{ if } |x|> p. \end{cases}$
\item[(ii)] $\overline{S}(x)=S(x^R)$ for each $x \in \dom S$.
\item[(iii)] $\big( S|\varphi \big) (x)=S(\varphi x)$ for each $\varphi x \in \dom S$.
\item[(iv)] $\big( \varphi |S \big) (x)=S(x\varphi)$ for each $x \varphi \in \dom S$.
\end{itemize}
\end{definition}

\begin{proposition} \label{Properties} Let $S$ be a tree, $p\geq 0$ an integer and $\varphi \in \zeta^*$: 
\begin{itemize}
\item[(i)] $\overline{\overline{S}}=S$;
\item[(ii)] $[\overline{S}]^p=\overline{[S]^p}$;
\item[(iii)] If $\depth(S)\leq 1$, then $\overline{S}=S$; 
\item[(iv)] $(S|\varphi)|\psi=S|(\varphi \psi)$,\,\,\,\, $\psi|(\varphi |S)=(\psi \varphi)|S$;
\item[(v)] $\overline{S | \varphi}=\varphi^R |\overline{S}$,\,\,\,\, $\overline{\varphi | S}=\overline{S} | \varphi^R$. 
\end{itemize}
\end{proposition}
\begin{proof} (i) is trivial. For (ii) we use $|x^R|=|x|$. (iii) If $|x|\leq 1$, then $x=x^R$, whereby $\overline{S}(x)=S(x^R)=S(x)$. (iv) $((S |\varphi)|\psi) (x)=(S|\varphi)(\psi x)=S(\varphi \psi x)=(S|(\varphi \psi))(x)$. The second equality in (iv) is similar. (v) $\overline{S|\varphi}(x)=S|\varphi(x^R)=S(\varphi x^R)=S(((\varphi x^R)^R)^R)=$ $S((x \varphi^R)^R)=\overline{S}( x \varphi^R)=(\varphi^R|\overline{S})(x)$. The second equality in (v) is similar.
\end{proof}

\begin{example} \label{ExampleOperators} Let us show some examples of the operators from Definition~\ref{Operators}. Let the sets be $\zeta=\{\alpha, \beta\}$, and $\Sigma=\{a, a', b, b', c,c'\}$. Figure~\ref{Fig2}(a) depicts the tree: 
\begin{align*}
S(x)=\begin{cases} a & \mbox{ if } x=1,\\
a' & \mbox{ if } x=\alpha,\\
b & \mbox{ if } x=\beta,\\
b' & \mbox{ if } x=\beta\alpha,\\
c & \mbox{ if } x=\beta\beta,\\
c' & \mbox{ if } x=\beta \beta\alpha.
\end{cases}
\end{align*}
Or, equivalently, $S$ is the set:
$$S=\{(1, a), (\alpha, a'), (\beta, b), (\beta \alpha, b'), (\beta^2, c), (\beta^2 \alpha, c')\}.$$
The operator $[S]^p$ takes the top of $S$ until depth $p$ and trims the rest of the tree. See, for example,  $[S]^1$ in Figure~\ref{Fig2}(b). The operator $S|\varphi$ takes the subtree of $S$ with the root at $\varphi$. For example, $S|\beta$ is the subtree depicted in Figure~\ref{Fig2}(c). 
The operator $\overline{S}$ consists in reversing the addresses:
$$\overline{S}=\{(1, a), (\alpha, a'), (\beta, b), ( \alpha \beta, b'), (\beta^2, c), ( \alpha \beta^2, c')\},$$
depicted in Figure~\ref{Fig2}(d). Finally, by the property Proposition~\ref{Properties}(v),  the operator $\alpha|S$ is the subtree with root at $\alpha^R=\alpha$ of the tree $\overline{S}$. 

Sometimes the operator $\overline{S}$ yields a tree whose representation can contains some gaps. Consider the following tree, with $\Sigma=\{a,b,c,d\}$ and $\zeta=\{\alpha, \beta, \gamma\}$:
$$S'=\{(1,a), (\alpha, b), (\beta, c), (\beta \gamma, d), (\beta^2, e)\}.$$
Figure~\ref{Fig2}(a') depicts $S'$, Figure~\ref{Fig2}(b') depicts $[S']^1$, Figure~\ref{Fig2}(c') depicts $S'|\beta$,  Figure~\ref{Fig2}(d') depicts $\overline{S'}$, and Figure~\ref{Fig2}(e') depicts $\gamma|S'$.  

\begin{figure}[tb] 
\centering
\includegraphics[height=53mm]{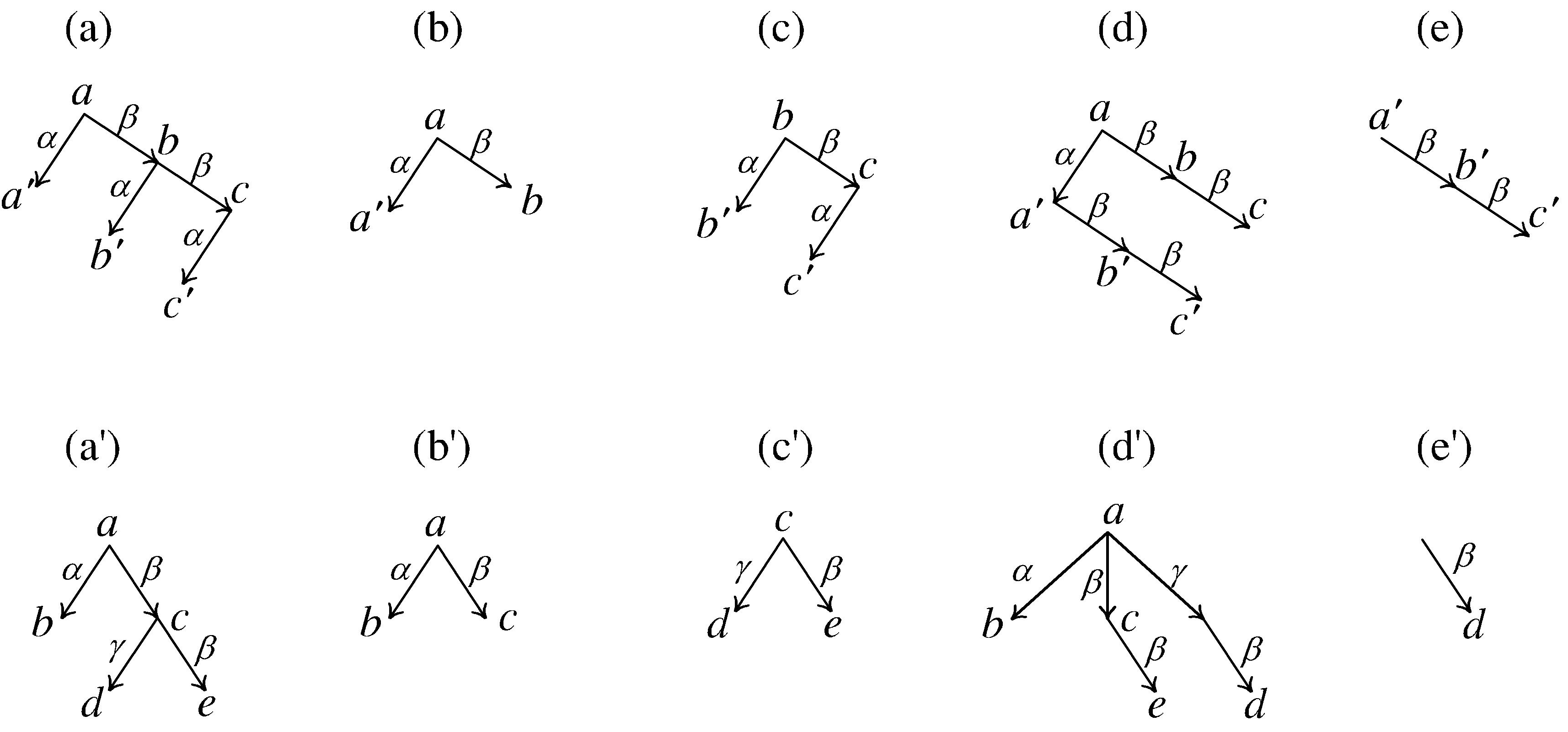}
\caption{Example of operators from Example~\ref{ExampleOperators}.}\label{Fig2}
\end{figure}
\end{example}

\begin{definition} We say that a string $x\in \Sigma^*$ is an \emph{ordering} of the tree $S \in \Tree$ if $|x|_a=|S^{-1}(a)|$ for each $a\in \Sigma$, where $|S^{-1}(a)|$ is the cardinality of the set $S^{-1}(a)$.

A \emph{linearisation} is a mapping $\Pi: \Tree_{\zeta, \Sigma} \longrightarrow \Sigma^*$ such that $\Pi(S)$ is an ordering of $S$, for each $S\in \Tree_{\zeta, \Sigma}$. 
\end{definition}

\begin{definition} A \emph{string language} is a subset of $\Sigma^*$ for some alphabet. A \emph{tree language} is a subset of $\Tree_{\zeta,\Sigma}$. By \emph{class} here we mean simply ``set'' and we reserve this denomination for \emph{class of string languages}, \emph{class of tree languages}, and \emph{class of linearisations}. 

Given a class of tree languages $\mathsf{X}$ and a class of linearisations $\mathsf{Y}$, we define the class of string languages:\footnote{More formally, $\frac{\mathsf{X}}{\mathsf{Y}}=\{ \Pi(W) \mid W \in \mathsf{X}, \Pi \in \mathsf{Y} \mbox{ such that } \Pi: \Tree_{\zeta, \Sigma}\longrightarrow \Sigma^*, W\subseteq \Tree_{\zeta, \Sigma} \}$.}
\begin{align}
\frac{\mathsf{X}}{\mathsf{Y}}=\{ \Pi(W) \mid W \in \mathsf{X}, \Pi \in \mathsf{Y}  \}.
\end{align}
\end{definition}

\section{Locality} \label{Locality} 

Locality is the mathematical notion according to which an object is identified by examining just a neighbourhood of its components.
An string language $L$ is said to be \emph{$p$-locally testable} \cite{zalcstein1972locally, yokomori1995polynomial}, $p$-local for short, if there is a set of prefix strings $U_1$, a set of internal strings $U_2$ and a set of suffix strings $U_3$ such that $x \in L$ if and only the prefix of $x$ of length $p$ is in $U_1$, every substring of $x$ of length $p$ is in $U_2$, and the suffix of $x$ of length $p$ is in $U_3$. A language is \emph{local} if it is $p$-local for some positive integer $p$. This notion is naturally translated to trees as described by Knuutila \cite{knuutila1993inference}. We adapt the definition to simple dependency trees.

\begin{definition} Let $S$ and $S'$ be trees and let $p$ be an integer $p\geq 0$. We say that:
\begin{itemize}
\item[(i)] $S'$ is the \emph{top $p$-subtree of} $S$ if $S'=[S]^p$.
\item[(ii)] $S'$ is a \emph{$p$-subtree of} $S$ if $[S|\varphi]^p=S'$ for some $\varphi \in \dom S$. 
\item[(iii)] $S'$ is a \emph{terminal $p$-subtree of} $S$ if $S|\varphi=S'$ for some $\varphi \in \dom S$ such that for all $x\in \zeta^*$ with $|x|>p$ we have that $\varphi x \not \in \dom S$.  
\end{itemize}
\end{definition}

\begin{definition} Given an integer $p>0$, we say that a tree language $W$ is \emph{$p$-local} if there exist three sets $U_1, U_2, U_3$ such that $S \in W$ if and only if the top $p$-subtree of $S$ is in $U_1$, every $p$-subtree of $S$ is in $U_2$ and every terminal $p$-subtree of $S$ is in $U_3$. We will write $\Loc(U_1,U_2,U_3)=W$. We say that a tree language is \emph{local} if it is $p$-local for some integer $p$. The class of local tree languages is denoted by $\LC$. 
\end{definition}

\begin{remark} It follows from the definition that if a local tree language $\Loc(U_1, U_2, U_3)$ is not empty, then $U_1, U_3 \subseteq U_2$. 
\end{remark}

\begin{example} Consider the set of trees $W_{\mathrm{squa}}=\{\emptyset, Q_a, Q_b, Q_{a,a}, Q_{a,b}, \ldots\}$ $\subseteq \Tree_{\zeta, \Sigma}$, where $\zeta=\{\alpha, \beta\}$ and $\Sigma=\{a,b\}$, displayed in Figure~\ref{Wsqua}. 
$W_{\mathrm{squa}}=\Loc (U_1,U_2,U_3)$, where:
\begin{itemize}
\item $U_1$ is the set of the trees of the form: 
$$\emptyset, \quad \includegraphics[height=11mm]{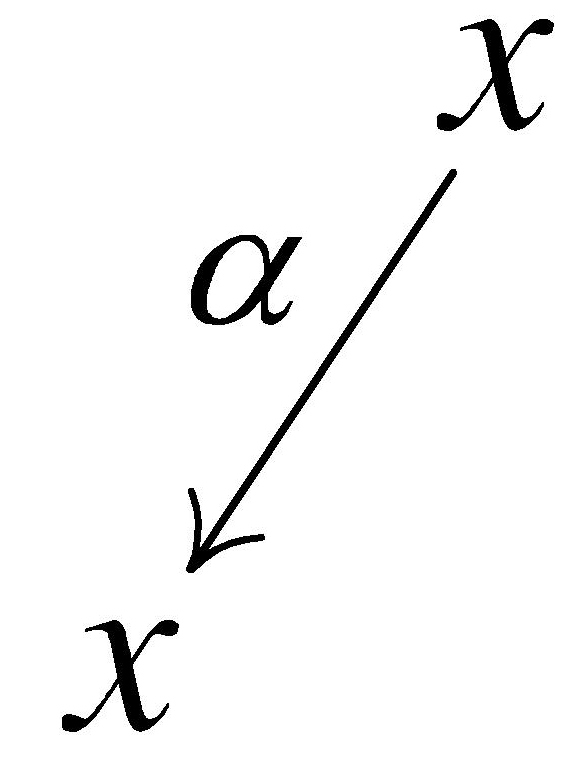}, \quad \includegraphics[height=16mm]{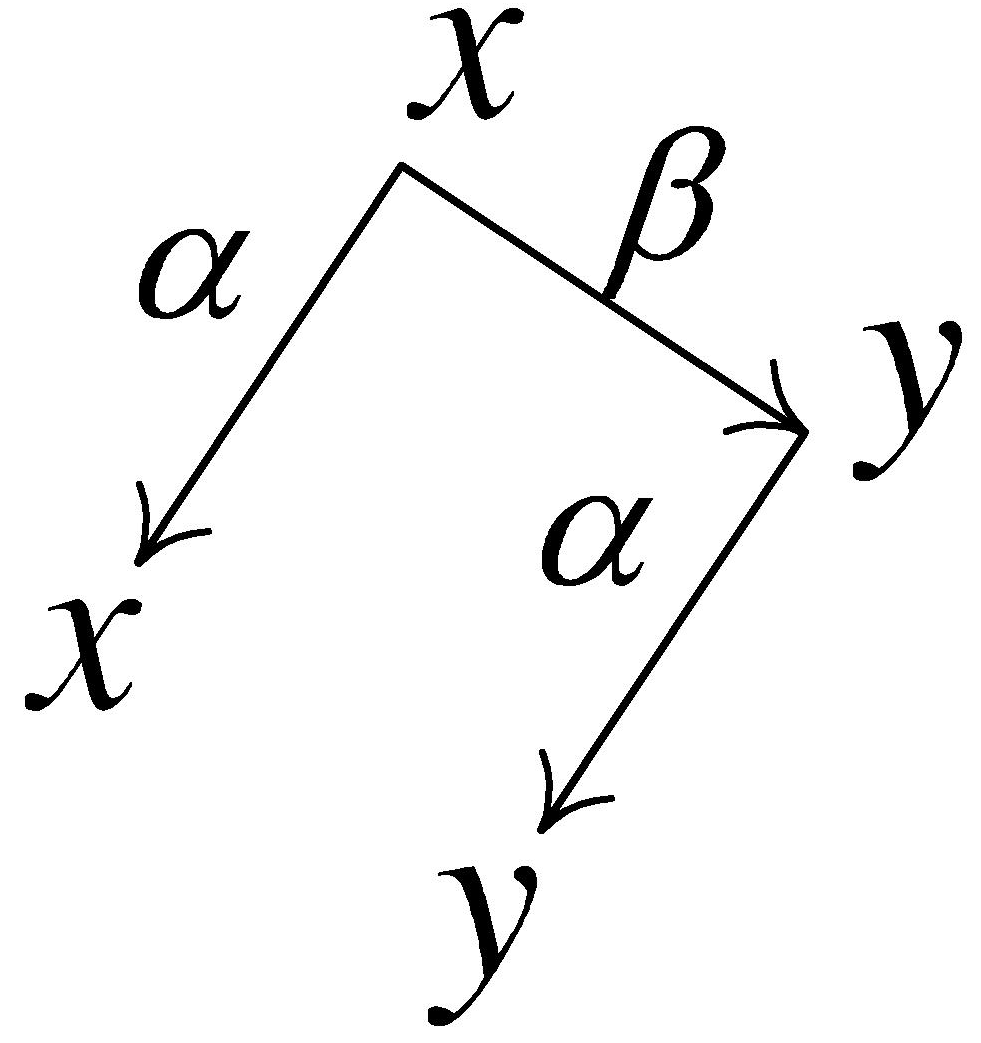}, 
\quad \includegraphics[height=16mm]{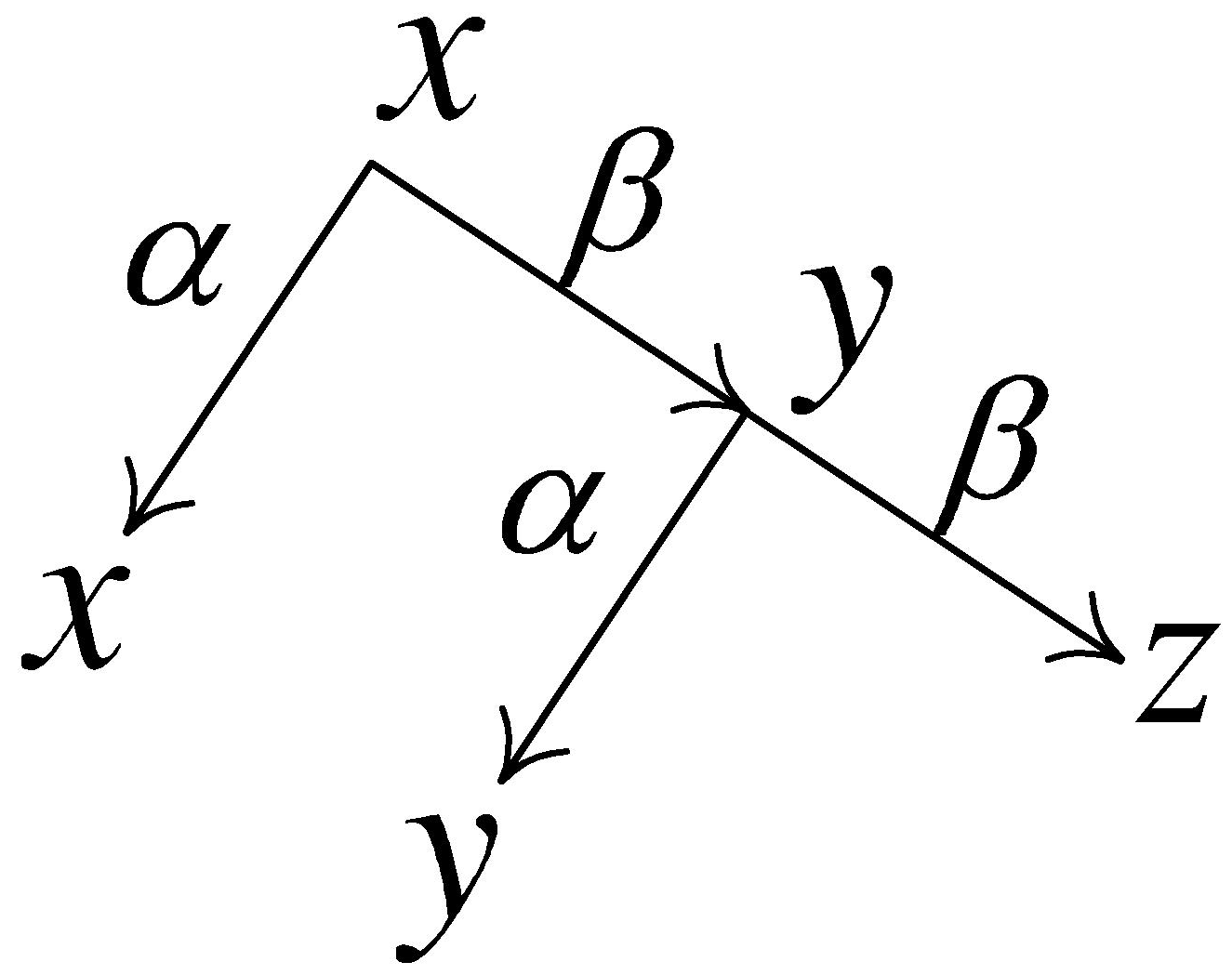},$$ 
with $ x,y,z \in \{a,b\}$.

\begin{figure} 
\centering
\includegraphics[height=26mm]{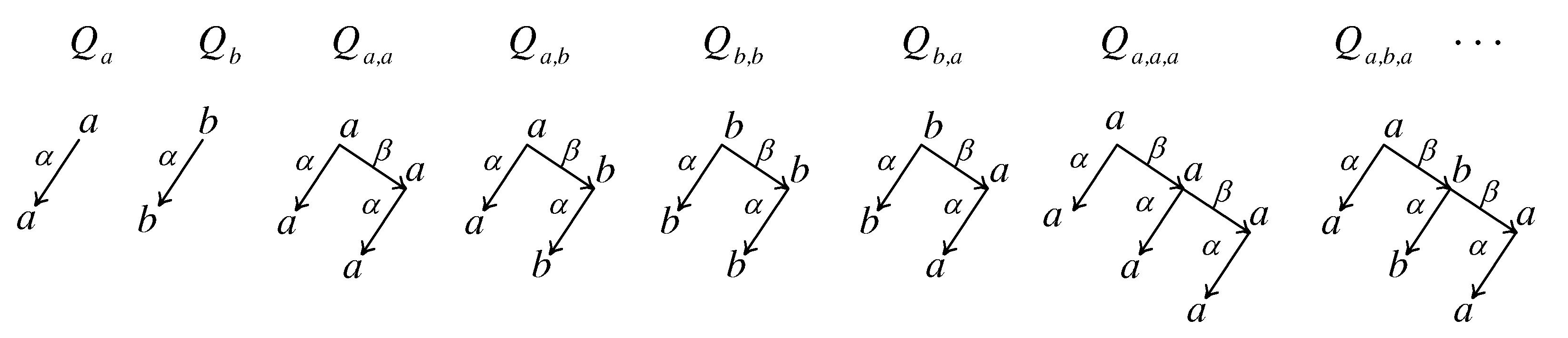}
\caption{The tree language $W_{\mathrm{squa}}$.}\label{Wsqua}
\end{figure}

\item $U_2=U_1$.

\item $U_3$ is the set of the trees of the form: 
$$\emptyset, \quad x^\bullet, \quad \includegraphics[height=11mm]{FigP1}, \quad \includegraphics[height=16mm]{FigP2},$$
with $ x,y,z \in \{a,b\}$.
\end{itemize}

\end{example}

\begin{example} Consider the set of trees $W_{\mathrm{mult}}=\{\emptyset, M_1, M_2, M_3$  $ \ldots\}$ $\subseteq \Tree_{\zeta, \Sigma}$, where $\zeta=\{\alpha, \beta, \gamma\}$ and $\Sigma=\{a,b,c\}$, displayed in Figure~\ref{Wmult}.
\begin{figure} 
\centering
\includegraphics[height=33mm]{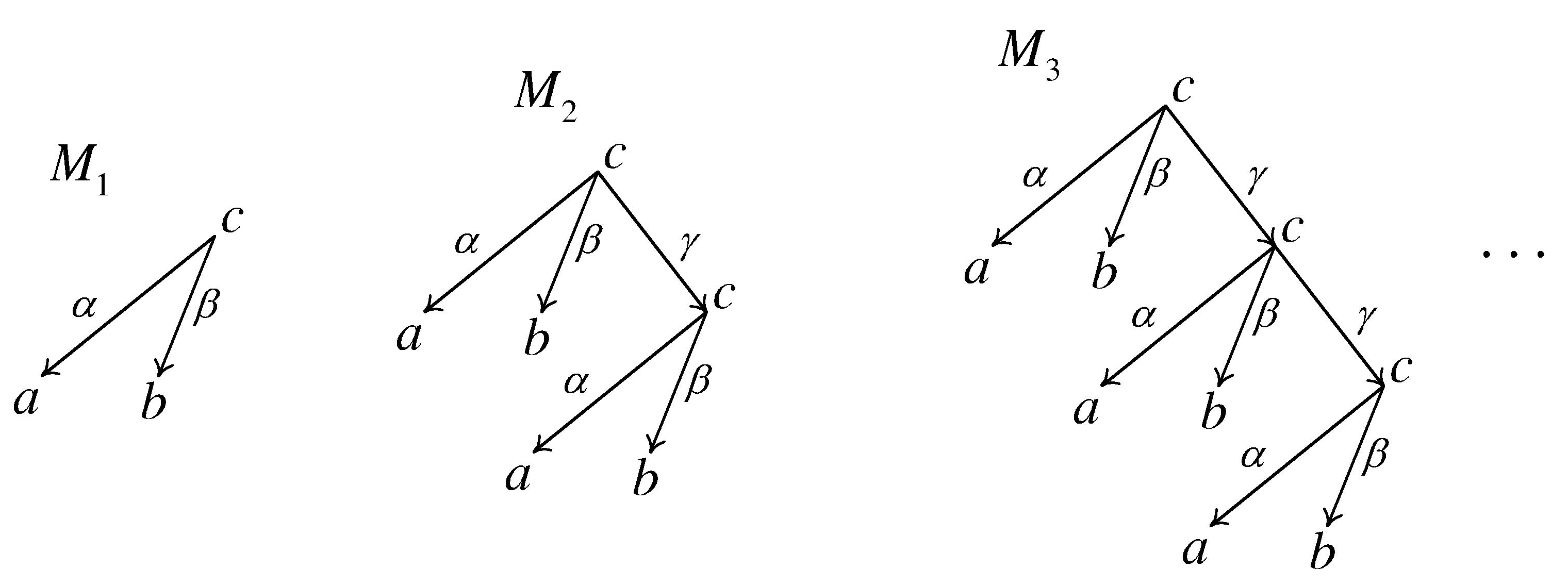}
\caption{The tree language $W_{\mathrm{mult}}$.}\label{Wmult}
\end{figure}
$W_{\mathrm{mult}}=\Loc(U_1,U_2, U_3)$, where:
\begin{itemize}
\item $U_1$ is the set of the trees:
$$ \emptyset, \qquad \includegraphics[height=14mm]{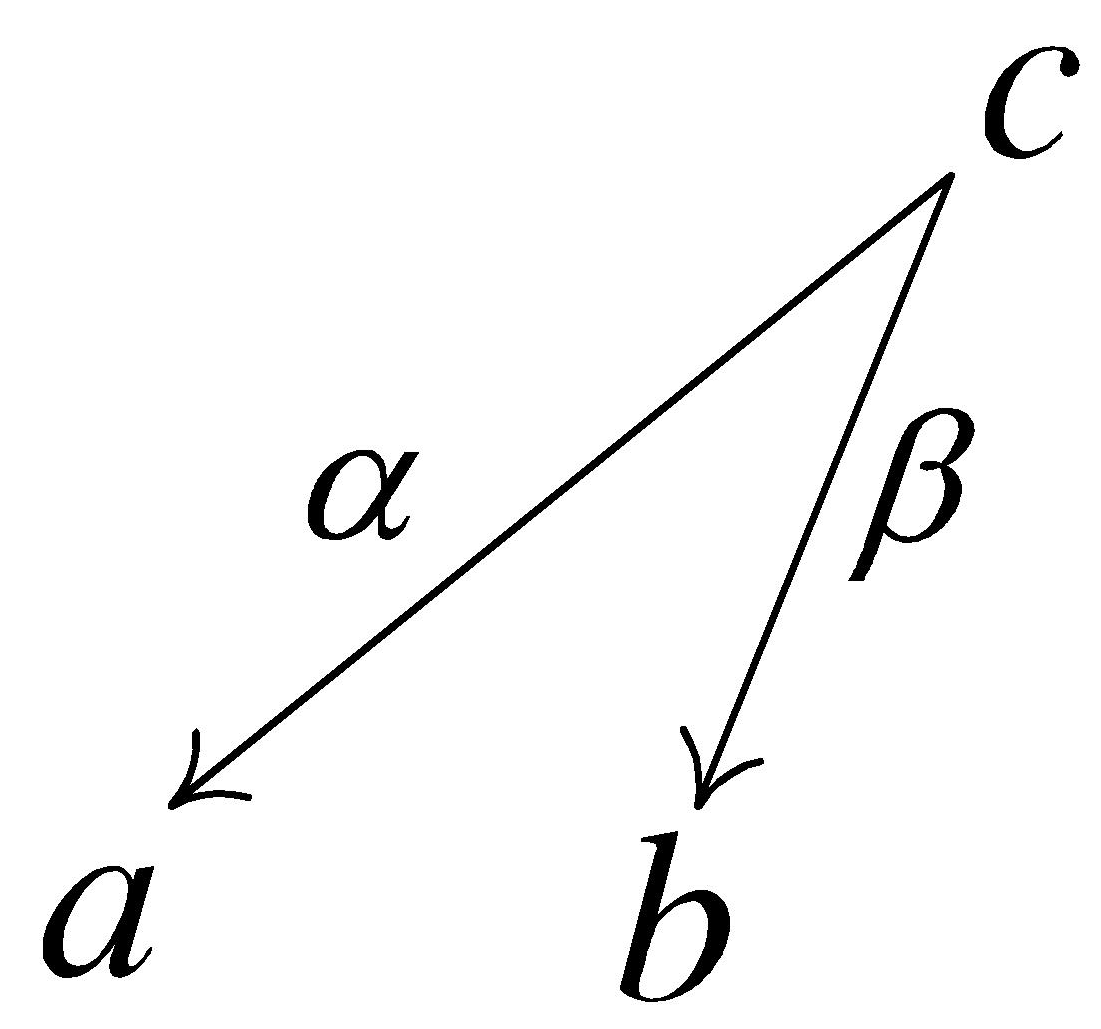}, 
\qquad \includegraphics[angle=0, height=14mm]{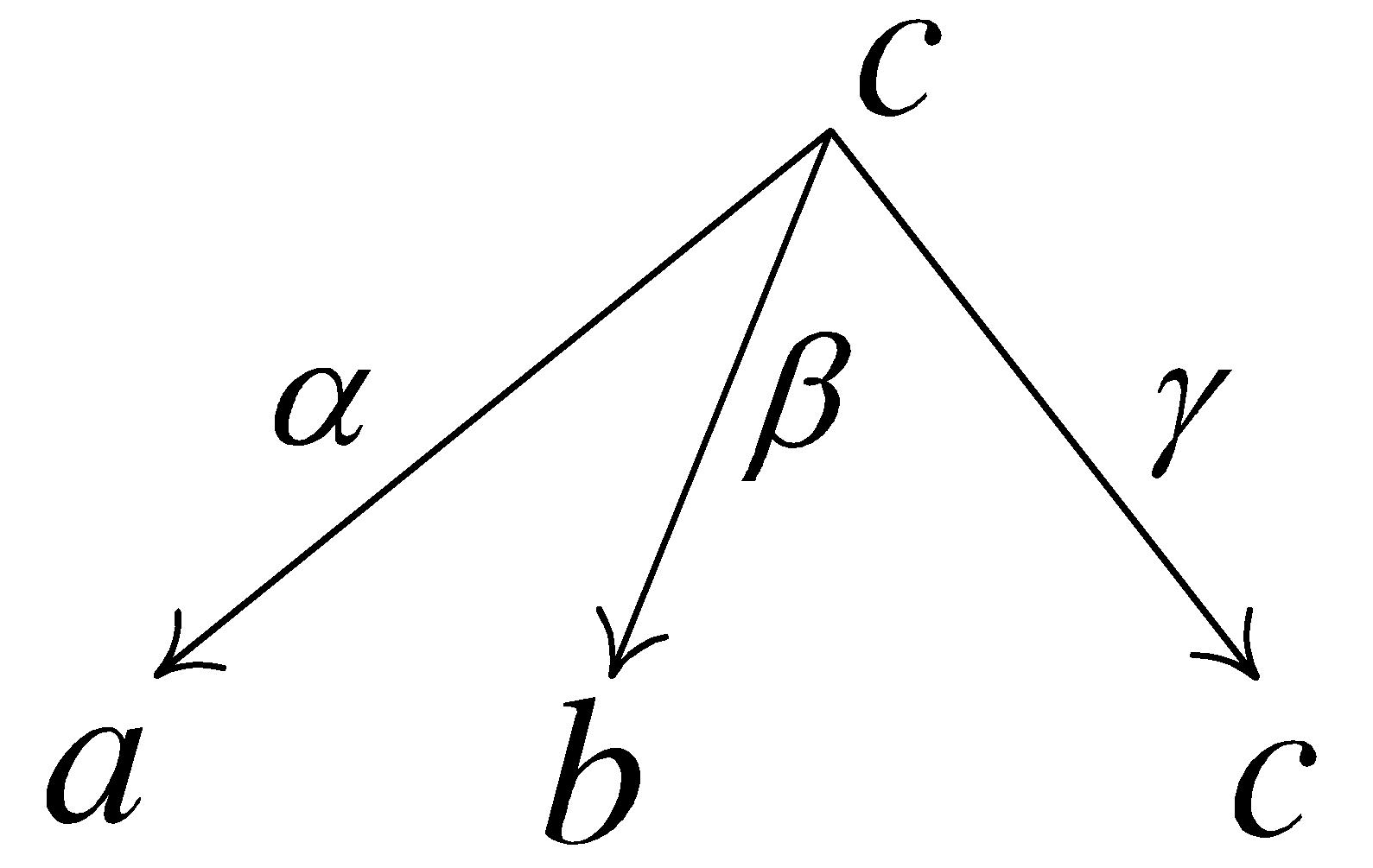} \quad.$$ 

\item $U_2=U_1$.

\item $U_3$ is the set of the trees:
$$ \emptyset, \qquad \includegraphics[height=14mm]{FigP4}.$$ 
\end{itemize}

\end{example}

\section{Projectivity} \label{Projectivity}

This section characterises the class of context-free languages as local trees projectively arranged, Theorem~\ref{ContextFreeCharac}. As described in Section~\ref{ProjectiveIntro}, a word-ordering of a given dependency tree is \emph{projective} if each subtree is a substring of the word-ordering. That is a broad notion, among other reasons, because the same tree can exhibit different projective orders in the same natural language. However, a definition that associates a single string with each tree will suffice for the mentioned characterisation.

\begin{definition} Let $\zeta=\{\alpha_1, \ldots, \alpha_m\}$ be a set of syntactic functions and $\Sigma$ a vocabulary. Fix a permutation $\rho$. We say that a mapping $\Pi: \Tree \longrightarrow \Sigma^*$ is a \emph{recursive projective linearisation} if for each $S\in \Tree$ we have:
\begin{align}
\Pi(\emptyset)&=\varepsilon,\\
\Pi(S)&=\rho(\Root(S), \Pi(S|\alpha_1), \ldots, \Pi(S|\alpha_m)),
\end{align}
where $\Root(S)=S(1)$ if $1 \in \dom S$, otherwise $\Root(S)=\varepsilon$. 
The class of such linearisations is denoted by $\PR$. 
\end{definition}

\begin{remark} Notice that the condition $\Pi(\emptyset)=\varepsilon$ stops the recursion. For an atomic tree $a^\bullet$ we have $\Pi(a^\bullet) =\Pi(\emptyset) \cdots\Pi(\emptyset)a\Pi(\emptyset)\cdots\Pi(\emptyset)=\varepsilon\cdots\varepsilon a\varepsilon\cdots\varepsilon=a.$

Notice, in addition, that $\Pi$ takes the root of $S$ and the strings $\Pi(S|\alpha_1),$ $ \ldots, \Pi(S|\alpha_m)$, and puts them in some order given by $\rho$. Since every $S|\alpha_i$ is a subtree, necessarily $\Pi$ transforms subtrees of $S$ into substrings of $\Pi(S)$. Therefore, $\Pi(S)$ is a projective ordering of $S$ (see Section~\ref{ProjectiveIntro}). 
\end{remark}

\begin{example} \label{ExampleWsquare} We consider the tree set $W_{\mathrm{squa}}$ and the recursive projective linearisation:
\begin{align}
\Pi_{\mathrm{squa}}(S)=\Pi_{\mathrm{squa}}(S|\alpha) \cdot \Root (S) \cdot \Pi_{\mathrm{squa}}(S|\beta).
\end{align}
Let us see an example of linearisation for a tree $Q_{a,b,a} \in W_{\mathrm{squa}}$. 
\begin{align*}
\Pi_{\mathrm{squa}}(Q_{a,b,a})&=\Pi_{\mathrm{squa}}(Q_{a,b,a}|\alpha)a\Pi_{\mathrm{squa}}(Q_{a,b,a}|\beta)\\
&=\Pi_{\mathrm{squa}}(a^\bullet)a\Pi_{\mathrm{squa}}(Q_{b,a})\\
&=aa\big(\Pi_{\mathrm{squa}}(Q_{b,a}|\alpha)b\Pi_{\mathrm{squa}}(Q_{b,a}|\beta)\big)\\
&=aa\big(\Pi_{\mathrm{squa}}(b^\bullet)b\Pi_{\mathrm{squa}}(Q_{a})\big)\\
&=aa (bb (\Pi_{\mathrm{squa}}(a^\bullet)a ))\\
&=aa (bb (aa ))\\
&=a^2b^2a^2.
\end{align*}
By generalizing this case to other trees in $W_\mathrm{squa}$, we obtain the square language $\Pi_{\mathrm{squa}}(W_\mathrm{squa})=L_\mathrm{squa}$. For the case of the tree language $W_{\mathrm{mult}}$ take the linearisation:
\begin{align}
\Pi_{\mathrm{mult}}(S)=\Pi_{\mathrm{mult}}(S|\alpha) \cdot\Pi_{\mathrm{mult}}(S|\beta) \cdot \Root(S) \cdot \Pi_{\mathrm{mult}}(S|\gamma).
\end{align}
We have, for example, for the tree $M_3$:
\begin{align*}
\Pi_{\mathrm{mult}}(M_3)&=\Pi_{\mathrm{mult}}(M_3|\alpha) \Pi_{\mathrm{mult}}(M_3|\beta) c \Pi_{\mathrm{mult}}(M_3|\gamma)\\
&=\Pi_{\mathrm{mult}}(a^\bullet) \Pi_{\mathrm{mult}}(b^\bullet) c \Pi_{\mathrm{mult}}(M_2)\\
&=abc\Pi_{\mathrm{mult}}(M_2)\\
&=abc\big(\Pi_{\mathrm{mult}}(M_2|\alpha) \Pi_{\mathrm{mult}}(M_2|\beta) c \Pi_{\mathrm{mult}}(M_2|\gamma)\big)\\
&=abc \big(\Pi_{\mathrm{mult}}(a^\bullet) \Pi_{\mathrm{mult}}(b^\bullet) c \Pi_{\mathrm{mult}}(M_1)\big)\\
&=abc\big(abc\Pi_{\mathrm{mult}}(M_1)\big)\\
&=abc\big(abc\big( \Pi_{\mathrm{mult}}(M_1|\alpha)\Pi_{\mathrm{mult}}(M_1|\beta)c\Pi_{\mathrm{mult}}(M_1|\gamma)  \big) \big)\\
&=abc\big(abc\big( \Pi_{\mathrm{mult}}(a^\bullet)\Pi_{\mathrm{mult}}(b^\bullet)c\Pi_{\mathrm{mult}}(\varepsilon^\bullet)  \big) \big)\\
&=abc(abc( abc \varepsilon ))\\
&=(abc)^3.
\end{align*}
Thus, $\Pi_{\mathrm{mult}}(W_\mathrm{mult})=L_\mathrm{mult}$. 
\end{example}

\begin{theorem}  \label{ContextFreeCharac} Let $\CF$ stand for the class of context-free languages: 
\begin{align} 
\frac{\LC}{\PR}= \CF.
\end{align}
\end{theorem}
\begin{proof}
$(\subseteq)$ Consider a $p$-local tree language $W=\Loc_p(U_1,U_2,U_3) \subseteq \Tree_{\zeta, \Sigma}$ and a projective recursive linearisation $\Pi$. We are going to prove that $\Pi(W)$ is a context-free language by constructing a context-free grammar $\mathscr{G}=(\Sigma, \mathcal{S}, \mathcal{V}, \mathcal{R})$, the language of which $\mathscr{L}(\mathscr{G})$ is $\Pi(W)$, where $\Sigma$ is the alphabet, $\mathcal{S}$ is the start symbol, $\mathcal{V}$ is the set of variables, and $\mathcal{R}$ is the set of rewriting rules.  

Let $\zeta=\{\alpha_1, \ldots, \alpha_m\}$ and let $\rho$ be the permutation associated to the linearisation $\Pi$. We can assume that $W\not= \emptyset$, otherwise we construct an empty context-free grammar. Recall that if $W=\Loc(U_1, U_2, U_3)$ is not empty, then $U_1, U_3 \subseteq U_2$. 
We take the set of variables, $\mathcal{V}=\{X_T \,|\, T \in U_2\}$, and we define the set of rules $\mathcal{R}$ as the union of the following sets:
\begin{itemize}
\item A set of start rules given by:
\begin{align} 
\mathcal{S}::=X_T
\end{align}
for each $T\in U_1$. 
\item A set of body rules:
\begin{align}
X_T::=\rho ( \Root(T), X_{T_1}, \ldots, X_{T_m})
\end{align}
for each set of trees $T,T_1, \ldots, T_m\in U_2$ such that $[T_i]^{p-1}=T|\alpha_i$ for each $i=1, \ldots, m$.
\item A set of terminal rules of the form:
\begin{align}
X_T::=\Pi(T)
\end{align}
for each $T \in U_3$.
\end{itemize}
With all this, we have that $\Pi(W)=\mathscr{L}(\mathscr{G})\in \mathsf{CF}$. 

$(\supseteq)$ For each context-free grammar there exists a grammar in Greibach normal form that generates the same language \cite{hopcroft1979ANTICintroduction, hopcroft1979introduction, blum1999greibach}. Consider the grammar in Greibach normal form with rules:
\begin{align}
\begin{cases} A_1::=a_1 B_{1,1}\cdots B_{r_1,1}\\
\,\,\, \vdots\\
A_s::=a_s B_{1,s}\cdots B_{r_s,s}
\end{cases}
\end{align}
where $A_j, B_{i,j} \in \mathcal{V}$  are variables and $a_j \in \Sigma$ are terminals symbols. We are going to construct a local set of trees that imitates the rules. Two trees can be chained if and only if the rules which represent invoke one to other. However, some of the variables $B_{i,j}$ can be repeated in the same rule. For this reason we introduce $r_1+\cdots+r_s$ new variables $C_{i,j}$, all different, that is, $C_{i,j}=C_{i',j'}$ if and only if $i=i', j=j'$. Consider the system:
\begin{align} \label{F15}
\begin{cases} A_1::=a_1 C_{1,1}\cdots C_{r_1,1}\\
\,\,\, \vdots\\
A_s::=a_s C_{1,s}\cdots C_{r_s,s}\\
C_{1,1}::=B_{1,1}\\
\,\,\, \vdots\\
C_{r_s,s}::=B_{r_s,s}\\
\end{cases}
\end{align}
This grammar generates the same language, and the rules does not contain repeated variables. Without loss of generality, we will assume a grammar $\mathscr{G}=(\Sigma, \mathcal{S}, \mathcal{V}, \mathcal{R})$ in the above form (\ref{F15}), where $\mathcal{V}=\{A_{j}\mid j=1,\ldots, s\}$ $\cup \{B_{i,j}\mid i=1,\ldots, r_j, \mbox{ and } j=1,\ldots, s\}$ $\cup \{C_{i,j}\mid i=1,\ldots, r_j, \mbox{ and } j=1,\ldots, s\}$, where all the variables $C_{i,j}$ are different. 
Next, we construct the set of subtrees, initial trees, and terminal trees, in order to define a local tree language. 
\begin{itemize}
\item  (Subtrees) Let $U_2$ be stand for the set of following trees. For each subset of $\mathcal{R}$ formed by $1+n+\sum_{j=1}^n k_j$ rules of the form:
\begin{align}
\begin{cases}
X::= a X_1\cdots X_n &\\
X_j::=b_j Y_{1,j}\cdots Y_{k_j,j} & j=1, \ldots, n\\
Y_{i,j}::=c_{i,j} Z_{i,j} & i=1, \ldots, k_j, j=1, \ldots, n
\end{cases}
\end{align}
where $X,X_i, Y_{1,i}\cdots Y_{n_i,i} \in \mathcal{V}$ and $a,b_i, c_{i,j} \in \Sigma\cup \{\varepsilon\} $, for each subscript and where $Z_{i,j} \in \mathcal{V}^*$, we construct a tree of depth 2 with $\Sigma$ as vocabulary and $\zeta=\mathcal{V}$ as syntactic functions as follows:
\begin{align}\label{treedepth2}
\includegraphics[height=30mm]{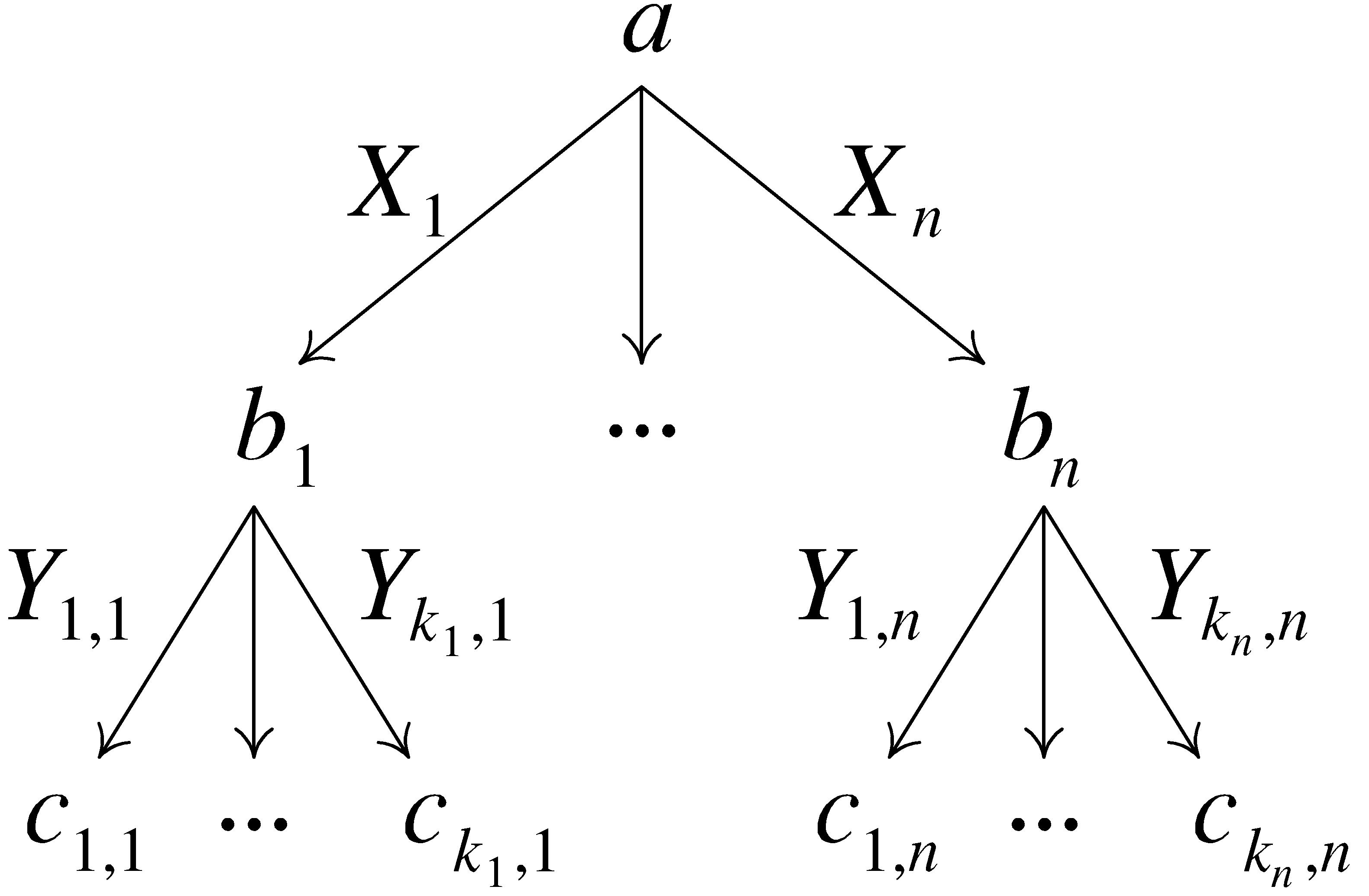}
\end{align}
For the superfluous rules of the form $C_{i,j}=B_{i,j}$, some letters $a, b_i, c_{i,j}$ can be the empty string. In this case, we do not define the tree at that Gorn adress.

\item (Initial trees) We do the same for the start rules. $U_1$ stands for all the following trees. 
	\begin{itemize} 
		\item For each set of rules of the form:
		\begin{align}
		\begin{cases}
		\mathcal{S}::= a X_1\cdots X_n &\\
		X_j::=b_j Y_{1,j}\cdots Y_{k_j,j} & j=1, \ldots, n\\
		Y_{i,j}::=c_{i,j} Z_{i,j} & i=1, \ldots, k_j, j=1, \ldots, n
		\end{cases}
		\end{align}
		we construct trees as in the case (\ref{treedepth2}). 
		\item For each set of rules of the form:
		\begin{align}
		\begin{cases}
		\mathcal{S}::= a X_1\cdots X_n &\\
		X_j::=b_j & j=1, \ldots, n\\
		\end{cases}
		\end{align}
		where $X,X_i, \in \mathcal{V}$ and $a,b_i \in \Sigma$, for each subscript, we 	construct a tree of depth 1 as follows:
		\begin{align}\label{treedepth1}
		\includegraphics[height=15mm]{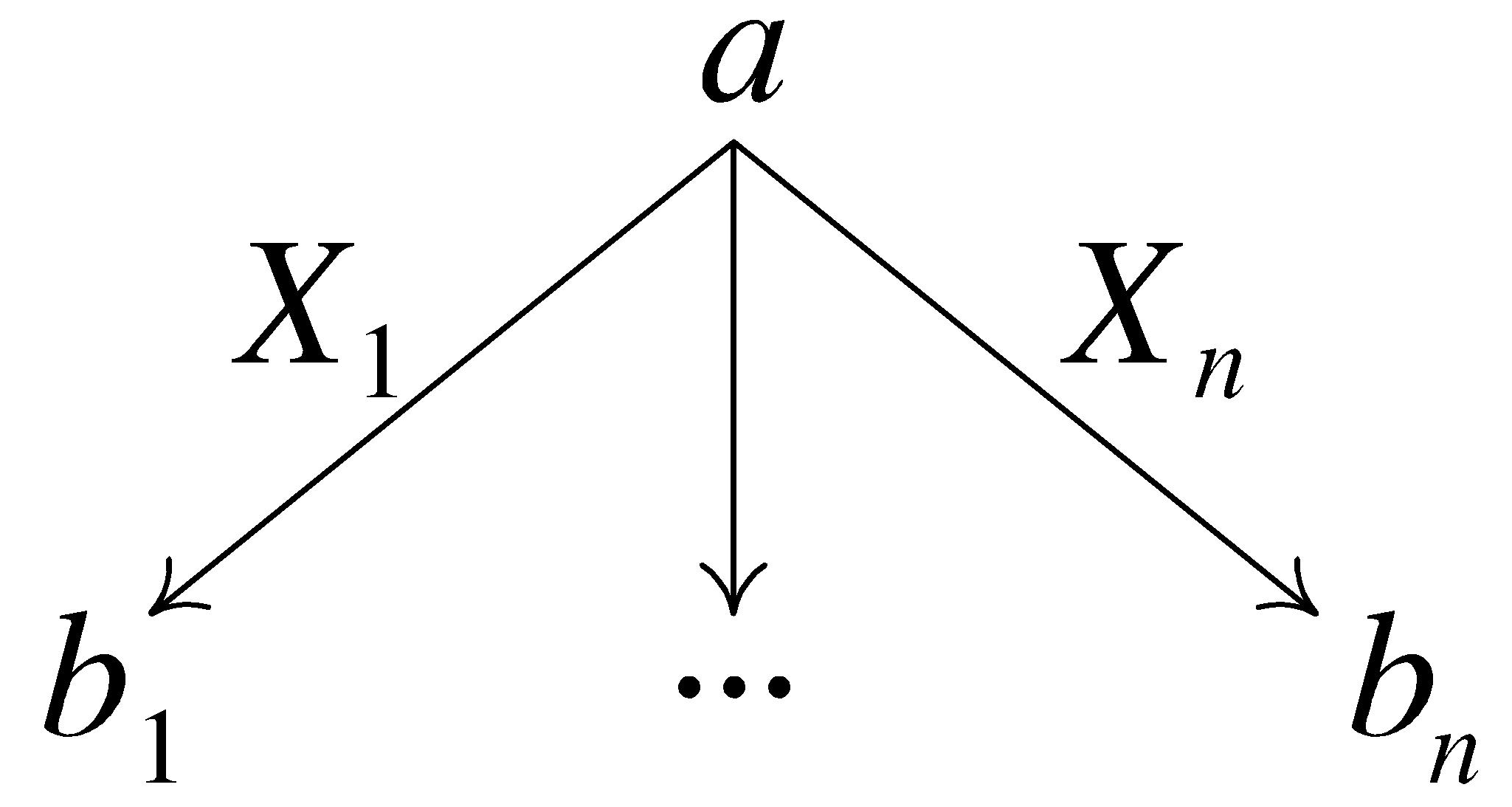}
		\end{align}
		\item For each rule of the form:
		\begin{align}
		\mathcal{S}&::= a
		\end{align}
		we construct an atomic tree $a^\bullet$. 
	\end{itemize}

\item (Terminal trees) Finally we construct trees for the terminal trees. $U_3$ stands for all the following trees.  

		\begin{itemize} 
		\item For each set of rules of the form:
		\begin{align}
		\begin{cases}
		X::= a X_1\cdots X_n &\\
		X_j::=b_j Y_{1,j}\cdots Y_{k_j,j} & j=1, \ldots, n\\
		Y_{i,j}::=c_{i,j}  & i=1, \ldots, k_j, j=1, \ldots, n
		\end{cases}
		\end{align}
		we construct a tree of the same form in the case (\ref{treedepth2}). 
		
		\item For each subset of rules of the form:
		\begin{align}
		\begin{cases}
		X::= a X_1\cdots X_n &\\
		X_j::=b_j & j=1, \ldots, n\\
		\end{cases}
		\end{align}
		we construct trees as in the case (\ref{treedepth1}). 
		
		\item For each rule of the form:
		\begin{align}
		X&::= a
		\end{align}
		where $X \in \mathcal{V}$ and $a \in \Sigma$, we construct an atomic tree 		$a^\bullet$.
		\end{itemize}
\end{itemize}

Next, define the recursive projective linearisation:
\begin{align}
\Pi(S)=\Root(S) \cdot \prod_{j=1}^{s} \prod_{i=1}^{r_j}\Pi(S|C_{i,j}) \cdot \prod_{j=1}^{s} \prod_{i=1}^{r_i}\Pi(S|B_{i,j})
\end{align}
which summarizes all the possibles orderings of each rewriting rule. In particular, notice that when the subtrees $S|C_{1,j}, \ldots, S|C_{r_j,j}$ are not empty for some $j$, all the other subtrees are empty. This yields:
\begin{align}
\Pi(S)=\Root(S) \cdot \varepsilon \cdots \varepsilon \cdot  \prod_{i=1}^{r_j}\Pi(S|C_{i,j}) \cdot \varepsilon \cdots \varepsilon= \Root(S)  \prod_{i=1}^{r_j}\Pi(S|C_{i,j}),
\end{align}
which reproduces the order of the rule $A_j::=a_{j} C_{1,j} \cdots C_{r_j,j}$. When the subtree $S|B_{i,j}$ is not empty for some $i$ and $j$, $\Root(S)=\varepsilon$ and the rest of subtrees are empty. This yields:
\begin{align}
\Pi(S)= \varepsilon \cdots \varepsilon \cdot \Pi(S|B_{i,j})  \cdot \varepsilon\cdots \varepsilon= \Pi(S|B_{i,j}),
\end{align}
which reproduces the order of the superfluous rule $C_{i,j}::=B_{i,j}$.
With all this,  we have that $\Pi(\Loc(U_1, U_2, U_3))=\mathscr{L}(\mathscr{G})$, which proves that $\CF\subseteq \LC/\PR$.
\end{proof}

\begin{remark} Notice that in the proof of Theorem~\ref{ContextFreeCharac} it is only required a depth two of locality.
\end{remark}

\section{Anti-classes} \label{AntiClasses}

We have used the operator $S|\varphi$ to extract subtrees of $S$ with the root at $\varphi$. Definitions of tree local languages and recursive projections use the subtree notion. Taking the alternative operator $\varphi | S$, we obtain dual definitions of locality and projectivity.

\begin{definition} Let $S$ and $S'$ be trees and let $p$ be an integer $p\geq 0$. 
\begin{itemize}
\item[(i)] We say that $S'$ is a \emph{$p$-anti-subtree of} $S$ if $[\varphi | S]^p=S'$ for some $\varphi \in \dom S$. 
\item[(ii)] We say that $S'$ is a \emph{terminal $p$-anti-subtree of} $S$ if $\varphi| S=S'$ for some $\varphi \in \dom S$ such that for all $x\in \zeta^*$ with $|x|>p$ we have that $x \varphi  \not \in \dom S$.  
\end{itemize}
\end{definition}

\begin{definition} Given an integer $p>0$, we say that a tree language $W$ is \emph{$p$-anti-local} if there are three sets $U_1, U_2, U_3$ such that $S \in W$ if and only if the top $p$-subtree of $S$ is in $U_1$, every $p$-anti-subtree of $S$ is in $U_2$, and every terminal $p$-anti-subtree of $S$ is in $U_3$. We write $\overline{\Loc}(U_1,U_2,U_3)=W$. A tree language is \emph{anti-local} if it is $p$-anti-local for some integer $p>0$. The class of anti-local tree languages is denoted by $-\LC$. 
\end{definition}

\begin{example} \label{ExampleAntiloc} Consider the following anti-local tree language $\overline{\Loc} (V_1,V_2,V_3)=\overline{W_{\mathrm{squa}}}$, depicted in Figure~\ref{Wantisqua}, where:
\begin{itemize}
\item $V_1$ is the set of the trees of the form:
$$\emptyset, \quad \includegraphics[height=11mm]{FigP1}, 
\quad \includegraphics[height=16mm]{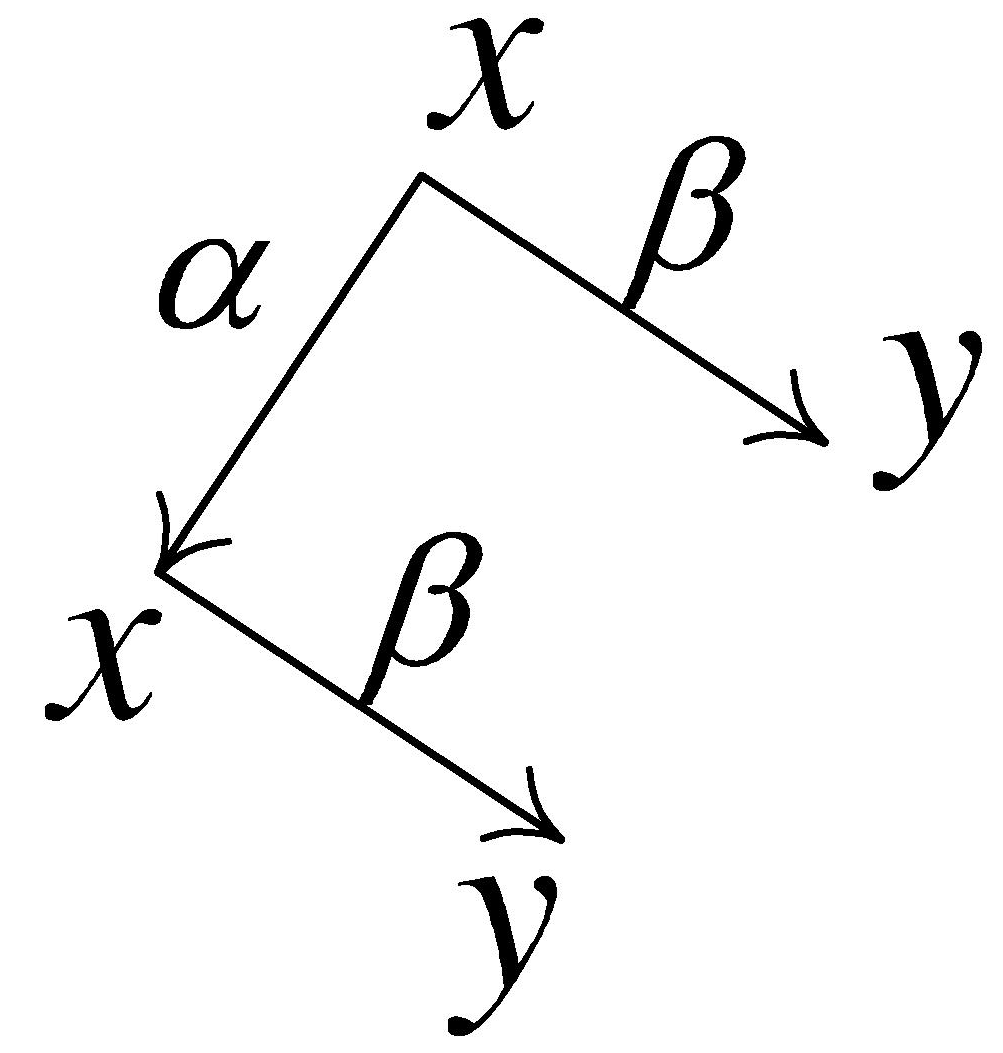}, 
\quad \includegraphics[height=16mm]{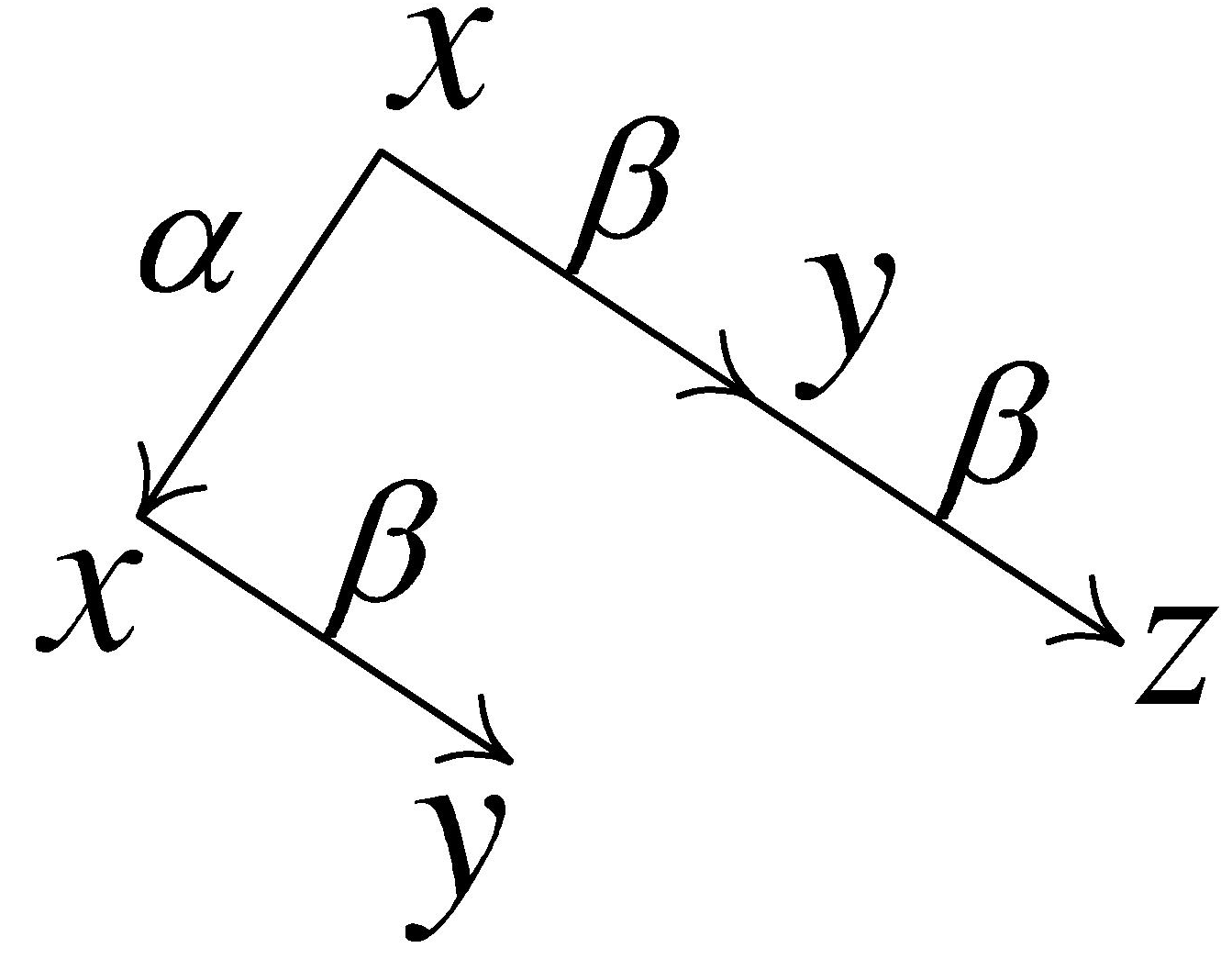},$$ 
with $ x,y,z \in \{a,b\}$.

\item $V_2=V_1$.

\item $V_3$ is the set of the trees of the form:
$$\emptyset, \quad x^\bullet, \quad \includegraphics[height=11mm]{FigP1}, 
\quad \includegraphics[height=16mm]{FigP8},$$
with $x,y,z \in \{a,b\}$.
\end{itemize}

\begin{figure} 
\centering
\includegraphics[height=26mm]{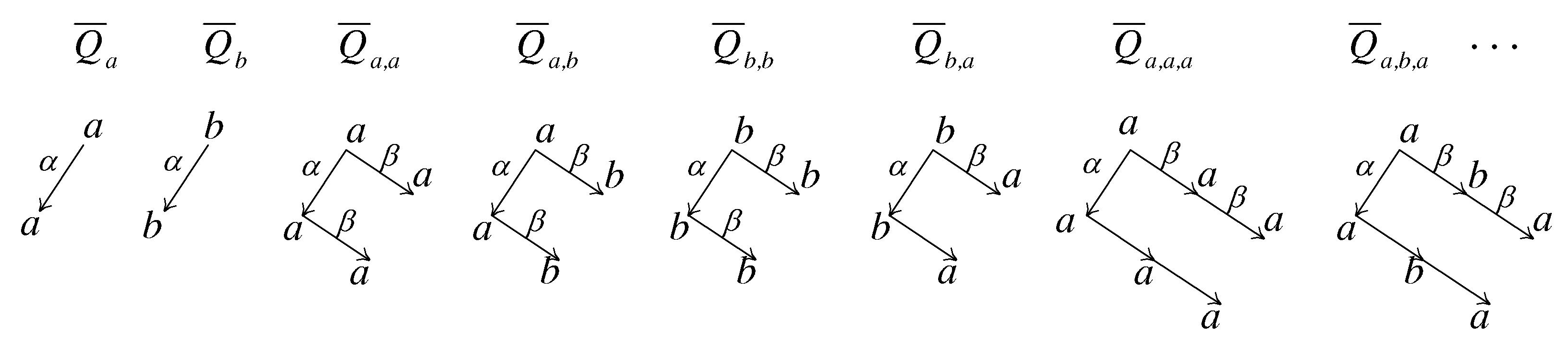}
\caption{The dual tree language $\overline{W}_{\mathrm{squa}}$ from the Example~\ref{ExampleAntiloc}.}\label{Wantisqua}
\end{figure}

\end{example}

\begin{definition} We say that a mapping $\overline{\Pi}: \Tree \longrightarrow \Sigma^*$ is a \emph{recursive anti-projective linearisation} if for each $S\in \Tree$ we have 
\begin{align}
\Pi(\emptyset)&=\varepsilon,\\
\Pi(S)&=\rho(\Root(S), \Pi(\alpha_1|S), \ldots, \Pi(\alpha_m|S)).
\end{align}
for some fixed permutation $\rho$. The class of such linearisations is denoted by $-\PR$. 
\end{definition}

\begin{lemma} \label{LemaGuai} We have the following commutations:
\begin{itemize}
\item[(i)] $\overline{\Loc (U_1,U_2, U_3)}=\overline{\Loc}(\overline{U_1},\overline{U_2},\overline{U_3})$; 
\item[(ii)] $\overline{\Pi}(S)=\Pi(\overline{S})$.
\end{itemize}
\end{lemma}
\begin{proof} (i) First, we see the following equivalences from Proposition~\ref{Properties}. 
\begin{itemize}
\item $S'$ is the top $p$-subtree of $S$ iff $\overline{S'}$ is the top $p$-subtree of $\overline{S}$. We just use the equivalence $S'=[S]^p$ iff $\overline{S'}=\overline{[S]^p}=[\overline{S}]^p$. 
\item $S'$ is a $p$-subtree of $S$ iff $\overline{S'}$ is a $p$-anti-subtree of $\overline{S}$. We just use the equivalence $S'=[S|\varphi]^p$ iff $\overline{S'}=\overline{[S|\varphi]^p}=[\overline{S|\varphi}]^p=[\varphi^R | \overline{S}]$ and the equivalence $\varphi \in \dom S$ iff $\varphi^R \in \dom \overline{S}$. 
\item $S'$ is a terminal $p$-subtree of $S$ iff $\overline{S'}$ is a terminal $p$-anti-subtree of $\overline{S}$. We use the equivalence $S'=S|\varphi$ iff $\overline{S'}=\overline{S|\varphi}=\varphi^R |\overline{S}$ and the equivalences $|x|>p$ iff $|x^R|>p$ and $\varphi x \not \in \dom S$ iff $x^R \varphi^R \not \in \dom \overline{S}$. 
\end{itemize}

Now we notice that: 
$$ S\in \overline{\Loc (U_1,U_2, U_3)} \iff  \overline{S} \in \Loc (U_1, U_2, U_3).$$
The right-hand expression is equivalent to say that:
\begin{itemize}
\item if $\overline{S'}$ is the top $p$-subtree of $\overline{S}$ then $\overline{S'} \in U_1$, or equivalently, if $S'$ is the top $p$-subtree of $S$, then $S' \in \overline{U_1}$. 
\item if $\overline{S'}$ is a $p$-subtree of $\overline{S}$, then $\overline{S'} \in U_2$, or equivalently, if $S'$ is a $p$-anti-subtree of $S$ then $S' \in \overline{U_2}$. 
\item if $\overline{S'}$ is a terminal $p$-subtree of $\overline{S}$, then $\overline{S'} \in U_2$, or equivalently, if $S'$ is a terminal $p$-anti-subtree of $S$, then $S' \in \overline{U_3}$. 
\end{itemize}
Therefore: 
$$S\in \overline{\Loc (U_1,U_2, U_3)} \iff  \overline{S} \in \Loc (U_1, U_2, U_3) \iff S \in \overline{\Loc}(\overline{U_1},\overline{U_2},\overline{U_3}).$$

(ii) By induction on the depth of the tree. If $\depth(S)\leq 1$, then $\overline{S}=S$, and trivially $\overline{\Pi}(S)=\Pi(S)=\Pi(\overline{S})$. 
Assume that the statement is true for any tree with depth less than $n$, and take a tree with $\depth(S)=n$. 
On the one hand, $\overline{S}(1)=S(1^R)=S(1)$. Thus, $\Root(S)=\Root(\overline{S})$. On the other hand, we notice that for any $\alpha_i \in \zeta$ we have $\depth(\alpha | S)<n$. Then, by hypothesis of induction:
$\overline{\Pi}(\alpha_i | S)=\Pi( \overline{ \alpha_i | S })= \Pi( \overline{S} | \alpha_i^R)=\Pi(\overline{S} | \alpha_i )$,
since $\alpha_i^R=\alpha_i \in \zeta$. Thus: 
\begin{align*}
\overline{\Pi}(S) &=\rho\big( \Root(S), \overline{\Pi}(\alpha_1 | S), \ldots, \overline{\Pi}(\alpha_m | S) \big) \\
&=\rho\big( \Root (S), \Pi(\overline{S}|\alpha_1 ), \ldots, \Pi(\overline{S}| \alpha_m ) \big) =\Pi(\overline{S}). 
\end{align*}
\end{proof}

\begin{theorem} \label{Teorema2}  The following equalities hold:
\begin{align}
\frac{\LC}{\PR}&=\frac{-\LC}{-\PR},\\
\frac{-\LC}{\PR}&=\frac{\LC}{-\PR}. 
\end{align}
\end{theorem}
\begin{proof} $L \in \frac{\LC}{\PR}$ if and only if $L=\Pi(\Loc(U_1,U_2, U_3))$ for some tree sets $U_1,U_2, U_3$. By Lemma~\ref{LemaGuai},
$L=\Pi(\Loc(U_1,U_2, U_3))= \overline{\Pi}(\overline{\Loc(U_1,U_2, U_3)})$ $=\overline{\Pi}(\overline{\Loc}(\overline{U_1}, \overline{U_2}, \overline{U_3})).$
Thus, $L \in \frac{\LC}{\PR}$ if and only if $L \in \frac{-\LC}{-\PR}$.
This proves the first equality. For the second equality, we suppose $L\in \frac{-\LC}{\PR}$. And then, 
$L=\Pi(\overline{\Loc}(U_1,U_2, U_3))= \Pi(\overline{\Loc(\overline{U_1},\overline{U_2},\overline{U_3})})$ $=\overline{\Pi}(\Loc (\overline{U_1},\overline{U_2},\overline{U_3}))$.
Thus, $L \in \frac{-\LC}{\PR}$ if and only if $L \in \frac{\LC}{-\PR}$.
\end{proof}

\begin{example} \label{Example25} We can see that $\Pi_{\mathrm{squa}}(\overline{W}_{\mathrm{squa}})=\Lcopy$. We write $T_{x_1, \ldots, x_n}$, with $x_1, \ldots, x_n \in \Sigma$ for the trees of the form: 
\begin{align*}
x_1\stackrel{\beta}\longrightarrow x_2 \stackrel{\beta}\longrightarrow \cdots \stackrel{\beta}\longrightarrow x_n
\end{align*}
Then, we have the following ordering for the tree $\overline{Q}_{\mathit{a,b,a}}$:
 \begin{align*}
\Pi_{\mathrm{squa}}(\overline{Q}_{a,b,a})&=\Pi_{\mathrm{squa}}(\overline{Q}_{a,b,a}|\alpha)a\Pi_{\mathrm{squa}}(\overline{Q}_{a,b,a}|\beta)\\
&=\Pi_{\mathrm{squa}}(T_{a,b,a})a\Pi_{\mathrm{squa}}(T_{b,a})\\
&=\big( \Pi_{\mathrm{squa}}(T_{a,b,a}|\alpha)a\Pi_{\mathrm{squa}}(T_{a,b,a}|\beta) \big)a\big( \Pi_{\mathrm{squa}}(T_{b,a}|\alpha)b\Pi_{\mathrm{squa}}(T_{b,a}|\beta) \big)\\
&=\big( \Pi_{\mathrm{squa}}(\emptyset)a\Pi_{\mathrm{squa}}(T_{b,a}) \big)a\big( \Pi_{\mathrm{squa}}(\emptyset) b\Pi_{\mathrm{squa}}(a^\bullet) \big)\\
&=\big( \varepsilon a \big( \Pi_{\mathrm{squa}}(T_{b,a}|\alpha) b  \Pi_{\mathrm{squa}}(T_{b,a}|\beta) \big) \big) a \big( \varepsilon ba \big)\\
&=\big( a \big( \Pi_{\mathrm{squa}}(\emptyset) b \Pi_{\mathrm{squa}}(a^\bullet) \big) \big) a\big(ba  \big) \\
&=\big( a \big(\varepsilon b a \big) \big) a\big(ba  \big) \\
&=( a (b a ) ) a (ba ) \\
&= (aba)^2. 
\end{align*}
Similarly, one can see that $\Pi_{\mathrm{mult}}(\overline{W}_{\mathrm{mult}})=\Lresp$.
\end{example}

\section{Anti-Context-Free Languages} \label{AntiContextFree}

Theorem~\ref{Teorema2} suggests a new class of languages: 
\begin{definition} We call 
\begin{align}
-\CF=\frac{-\LC}{\PR}=\frac{\LC}{-\PR}
\end{align}
the class of \emph{anti-context-free languages}. Given two languages $L \in \CF$ and $L' \in -\CF$, we say that they form a \emph{dual pair}, written $L \perp L'$, if $L=\Pi(W)$ and $L'=\overline{\Pi}(W)=\Pi(\overline{W})$ for some local tree language $W$ and some projective linearisation $\Pi$. A language $L$ is self-dual if $L \perp L$. 
\end{definition}

\begin{remark} The relation $\perp$ is symmetric but not in general reflexive nor transitive. 
\end{remark}

\begin{example} \label{exampperp} We know that $W_{\mathrm{squa}} \in \LC$ and $\Pi_{\mathrm{squa}} \in \PR$. Since $\Pi_{\mathrm{squa}}(\overline{W_{\mathrm{squa}}})=\Lcopy$, we have that $\Lcopy \in -\CF$. Similarly, we have that $\Lresp \in -\CF$. Thus, we have the dual pairs $\Lsqua \perp \Lcopy$ and $\Lmult \perp \Lresp$.
\end{example}

\begin{example} \label{multiduality} A language does not have a unique dual language. This is because for a language $L \in \CF$ it is possible to find different $W,W',\Pi,\Pi'$ such that $\Pi(W)=\Pi'(W')=L$. Let us examine the following case. 
First we consider another classical language called the \emph{Dyck language}, $L_{\mathrm{Dyck}} \subseteq (\Sigma \cup \widetilde{\Sigma})^*$, where $\Sigma=\{\,\, (, [, \{, [\![ \, \,\}$ is the alphabet of left parenthesis and  $\widetilde{\Sigma}=\{ \,\, ), ], \}, ]\!]\,\, \}$ the alphabet of right parenthesis. $L_{\mathrm{Dyck}}$ consists in the strings of \emph{well balanced parentheses} \cite{berstel2013transductions}. This language is representative of context-freeness in the sense stated in the Chomsky-Sch\"utzenberger Theorem \cite{chomsky1963algebraic, berstel2013transductions, autebert1997context}. 

Consider a pair of languages very similar to $\Lsqua$ and $\Lcopy$. Define the operator $\widetilde{(\cdot)}:\Sigma \longrightarrow \widetilde{\Sigma}$ by $\tilde{(}={)}$, $\tilde{[}={]}$, $\tilde{\{}={\}}$ and $\tilde{[\![}={]\!]}$. With this, we define:
\begin{align}
\widetilde{\Lsqua} &=\{ a_1 \tilde{a}_1 \cdots a_n\tilde{a}_n \,|\, a_1,\ldots, a_n \in \Sigma, n\geq 0\}\\
\widetilde{\Lcopy} &=\{ a_1 \cdots a_n \tilde{a}_1 \cdots \tilde{a}_n \,|\, a_1,\ldots, a_n \in \Sigma, n \geq 0\}
\end{align}
These languages are, respectively, context-free (indeed regular) and non-context-free. 

For the Dyck language, we can construct a local tree language for which the following Figure~\ref{FigExampleMultDual}(a) shows a tree $S$. The function $\alpha$ opens a parenthesis, $\gamma$ closes the parenthesis, $\beta$ puts material inside the parenthesis, while $\delta$ puts material at the right. We take the linearisation: 
\begin{align} 
\Pi(S)=\Root(S) \cdot \Pi(S | \alpha) \cdot \Pi(S | \beta) \cdot \Pi(S | \gamma) \cdot \Pi(S | \delta)
\end{align}
For example, for the tree of Figure~\ref{FigExampleMultDual}(a)  we have the string: 
$\Pi(S)=[\,(\,)\,]\,\{\,[\![\,]\!]\,\}$. Figure~\ref{FigExampleMultDual}(b) depicts the reversed tree $\overline{S}$ which, when it is linearized, yields the string $\Pi(\overline{S})=[\, (\, [\![ \, \{ \, ] \, ) \, ]\!] \, \} \in \widetilde{\Lcopy}$. 
And in general we have that $\LDyck \perp \widetilde{\Lcopy}$.

\begin{figure}[tb] 
\centering
\includegraphics[height=72mm]{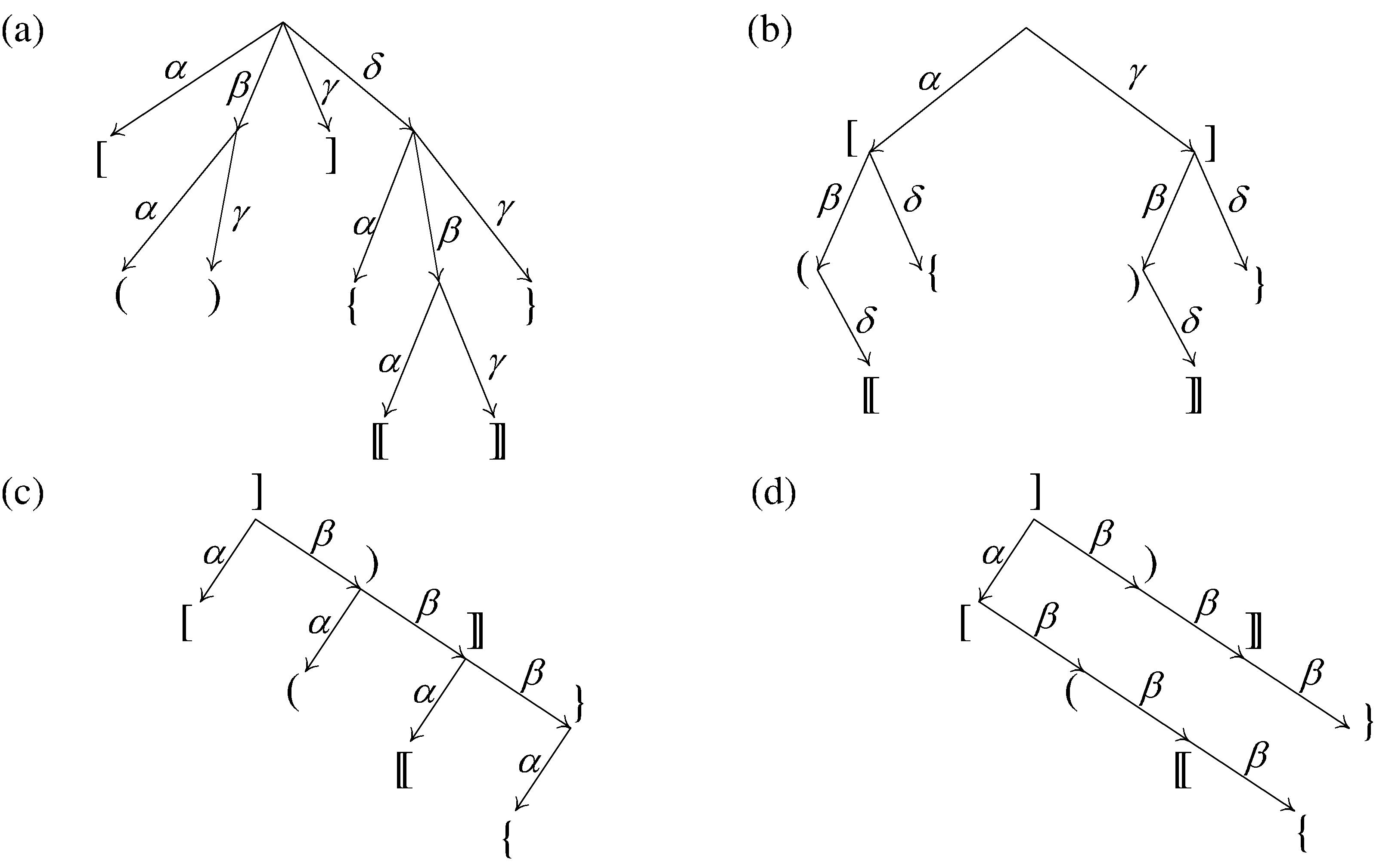}
\caption{Trees from Example~\ref{multiduality}.}\label{FigExampleMultDual}
\end{figure}

 Now we consider the following Figures. Figure~\ref{FigExampleMultDual}(c) depicts a tree for a tree language defined locally with which the linearisation:
\begin{align}
\Pi(S)=\Pi(S | \alpha)  \cdot \Root(S) \cdot \Pi(S | \beta)
\end{align}
yields the language  $\widetilde{\Lsqua}$. In particular the tree from the figure yields the string $[\,]\,(\,)\,[\![ \, ]\!]\, \{\,\}$.  Figure~\ref{FigExampleMultDual}(d) shows the reversed tree with which the same linearisation yields the string  $[ \, ( \, [\![ \, \{ \, ] \, ) \, ]\!] \, \} \, \in  \widetilde{\Lcopy}$.
In sum, $\widetilde{\Lcopy} \perp \LDyck $ and $  \widetilde{\Lcopy}  \bot \widetilde{\Lsqua}$, whereby a language can have more than one dual language.  
\end{example}

Let us see some basic properties of anti-context-free languages. We write $\UN$ for the class of unary languages, that is, languages defined over an alphabet with only one letter; $\FN$, for the class of finite languages; $\RG$ for the class of regular languages; and $\LS$ the class of local string languages. 
\begin{definition}
Given a subclass $\mathsf{X} \subseteq \CF \cup -\CF$ we define the `\emph{anti}' operator: 
\begin{align}
-\mathsf{X}=\{ L' \in \CF \cup -\CF \mid L' \perp L, L \in \mathsf{X}\}.
\end{align}
We say that a class $\mathsf{X} \subseteq \CF \cup -\CF$ is involutive if and only $ -(-\mathsf{X})=\mathsf{X}$.
\end{definition}

It is easily seen that $\mathsf{X} \subseteq -(-\mathsf{X})$, since $\perp$ is a symmetric relation, but in general, the `anti' operator is not involutive. $\CF$ and $-\CF$ are trivially involutive classes. For some classes, we have that $-\mathsf{X}=\mathsf{X}$, and then the class is trivially involutive.

\color{black}
\begin{proposition} We have the following properties for the `anti' operator:
\begin{itemize}
\item[(i)] $-\CF \not=\CF$;
\item[(ii)] $-(\CF \cap \UN) = -(\RG \cap \UN)= \RG \cap \UN = \CF \cap \UN$;
\item[(iii)] $-\FN = \FN$;
\item[(iv)] $\RG$ is not involutive. 
\item[(v)] $L=\{a^nb^n \mid n\geq 0\}$ is self-dual. In particular, $L \in \CF \cap -\CF$. 
\item[(vi)] Every local string language is self-dual. In particular, $\LS \subseteq -\LS$.  
\item[(vii)] $L_{\mathrm{mult}}$ is self-dual. 
\item[(viii)] $\LS$ is not involutive. 
\item[(ix)] $\CF \cap -\CF$ is not involutive. 
\end{itemize}
\end{proposition} 
\begin{proof}
\begin{itemize}
\item[(i)] By Example~\ref{exampperp}, $\Lsqua \perp \Lcopy$. Since $\Lcopy \not \in \CF$, the classes must be different. 
\item[(ii)] It is easily seen that $\RG \cap \UN = \CF \cap \UN$. Consider a unary language $L=\Pi(W)$ in $\CF$. Since $L$ is unary, for any linearisation $\Pi$, if $S\in W$, then $\Pi(S)=\overline{\Pi}(S)$. And, then $L=\Pi(W)=\overline{\Pi}(\overline{W})=\Pi(\overline{W})$. Hence, for any unary language, $L \in \CF$ if and only if $L \in -\CF$. 
\item[(iii)] $\Pi(W)$ is a finite language if and only if $W$ is finite if and only if $\overline{\Pi}(W)$ is finite. 
\item[(iv)] recall from the Example~\ref{multiduality} that $\widetilde{\Lsqua} \perp \widetilde{\Lcopy}$ and $\widetilde{\Lcopy} \perp \LDyck $. $\widetilde{\Lsqua}$ is regular, $\LDyck $ is context free but not regular. Thus, $\mathsf{RG} \subsetneq -(-\mathsf{RG})$.
\item[(v)] Let $W$ be the local tree language:
$$\includegraphics[height=24mm]{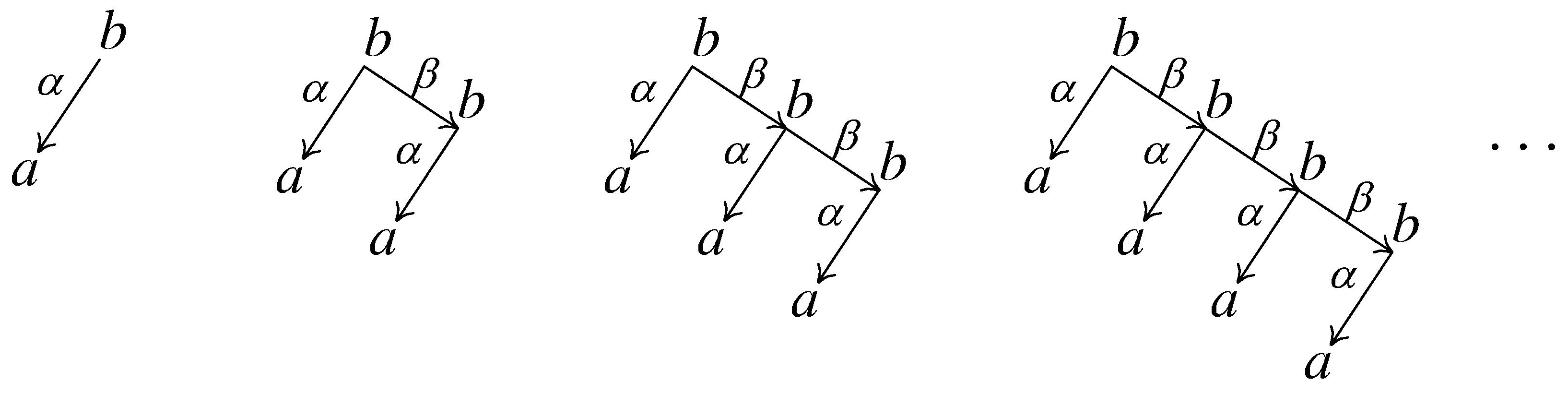}$$
\noindent By taking the linearisation $\Pi(S)=\Pi(S|\alpha) \cdot \Pi(S|\beta)\cdot \Root(S)$ we have that $L=\{a^nb^n \mid n \geq 0 \}=\Pi(W)=\overline{\Pi}(W)$, and therefore, $L\perp L$, and $L \in \CF \cap -\CF$. 
\item[(vi)] A string $x=x_1x_2 \cdots x_n$, with $x_1, x_2, \ldots, x_n \in \Sigma$ can be encoded as a tree with only one syntactic function $\alpha$:
\begin{align*}
 T_{x}=\quad x_1\stackrel{\alpha}\longrightarrow x_2 \stackrel{\alpha}\longrightarrow \cdots \stackrel{\alpha}\longrightarrow x_n
\end{align*}
A local string languages $L$ is defined by a set of prefixes, a set of strings and a set of sufixes. Encoding these strings we obtain a set of top trees, a set of subtrees, and a set of terminal subtrees that give us a local tree language $W_L$. When we linearize the trees by 
$\Pi(S)=\Root(S) \cdot \Pi(S|\alpha)$, we obtain the local string language in question, $\Pi(W_L)=L$. 
Now we only need to notice that $\overline{T_x}=T_x$. This means that $L=\Pi(W_L)=\Pi(\overline{W_L})$. Therefore, $L\perp L$. 
\item[(vii)] We take the set of prefix strings $\{a, ab, abc\}$, the set of internal strings $\{abc\}$ and the set of suffix strings $\{abc, bc, c\}$. The local string language defined by these sets is  $L_{\mathrm{mult}}$. By (vi) the language is self-dual. 
\item[(viii)] By (vii) and (vi), $L_{\mathrm{mult}}$ is self-dual. However, we saw that $\Lmult \perp \Lresp$. But $\Lresp$ is not context-free, neither a local string language. 
\item[(ix)] The same reasoning of (viii). 
\end{itemize}
\end{proof}

We set $\Sigma=\{a_1, \ldots, a_k\}$. The \emph{Parikh mapping} is the mapping $p:\Sigma^* \longrightarrow \mathbb{N}^k$ defined by  $p(x)=(|x|_{a_1}, \ldots, |x|_{a_k})$. A set in $\mathbb{N}^k$ is \emph{semi-linear} if it is a finite union of sets of the form $\{ p_1 A_1+ \cdots +p_n A_n +B \,|\, p_1, \ldots, p_n \in \mathbb{N}\}$ for some $n\geq 0$ and $A_1, \ldots, A_n, B \in \mathbb{N}^k$. A string language $L\subseteq \Sigma^*$ is semi-linear if $p(L)$ is a semi-linear set   \cite{kallmeyer2010parsing}. The Parikh theorem states that context-free languages are semi-linear \cite{parikh1966context}. 

\begin{proposition} Languages in $-\CF$ are semilinear. 
\end{proposition}
\begin{proof} First we notice that the notion of semi-linear string language can be directly translated to tree languages. We define the Parikh mapping for trees $P:\Tree_{\zeta, \Sigma} \longrightarrow \mathbb{N}^k$ as $P(S)=(|S^{-1}(a_1)|, \ldots, |S^{-1}(a_k)|)$. A tree language $W\subseteq \Tree_{\zeta, \Sigma}$ is said to be semi-linear if $P(W)$ is a semi-linear set.

Let $W$ be a tree language and let $\Phi: \Tree_{\zeta, \Sigma} \longrightarrow \Sigma^*$ be a linearisation. We have that $\Phi(W)$ is a semi-linear string language if and only if $W$ is a semi-linear tree language. This is because $P(S)=p(\Phi(S))$.

Finally, we consider $L'\in -\CF$. There is a $L \in \CF$ such that $L\perp L'$. That is, there is a projective linearisation $\Pi$ and a local tree language $W$ such that $\Pi(W)=L$ and $\overline{\Pi}(W)=L'$. By the above comment, $L'$ is semi-linear if and only if $W$ is semi-linear if and only if $L$ is semi-linear. By Parikh's theorem, $L$ is semi-linear, so $L'$ is semi-linear.
\end{proof}

\section{Discussion} \label{Discussion}

Since context-free languages seem to form a fundamental class for understanding the human language, although insufficient in terms of weak capacity, the early strategy slightly increased the descriptive power of the context-free grammars. Although this ancient recipe may be formally appropriate, it ignores the phenomenon of duality, explained here, which has, we believe, linguistic relevance.

\begin{figure} 
\centering
\includegraphics[height=61mm]{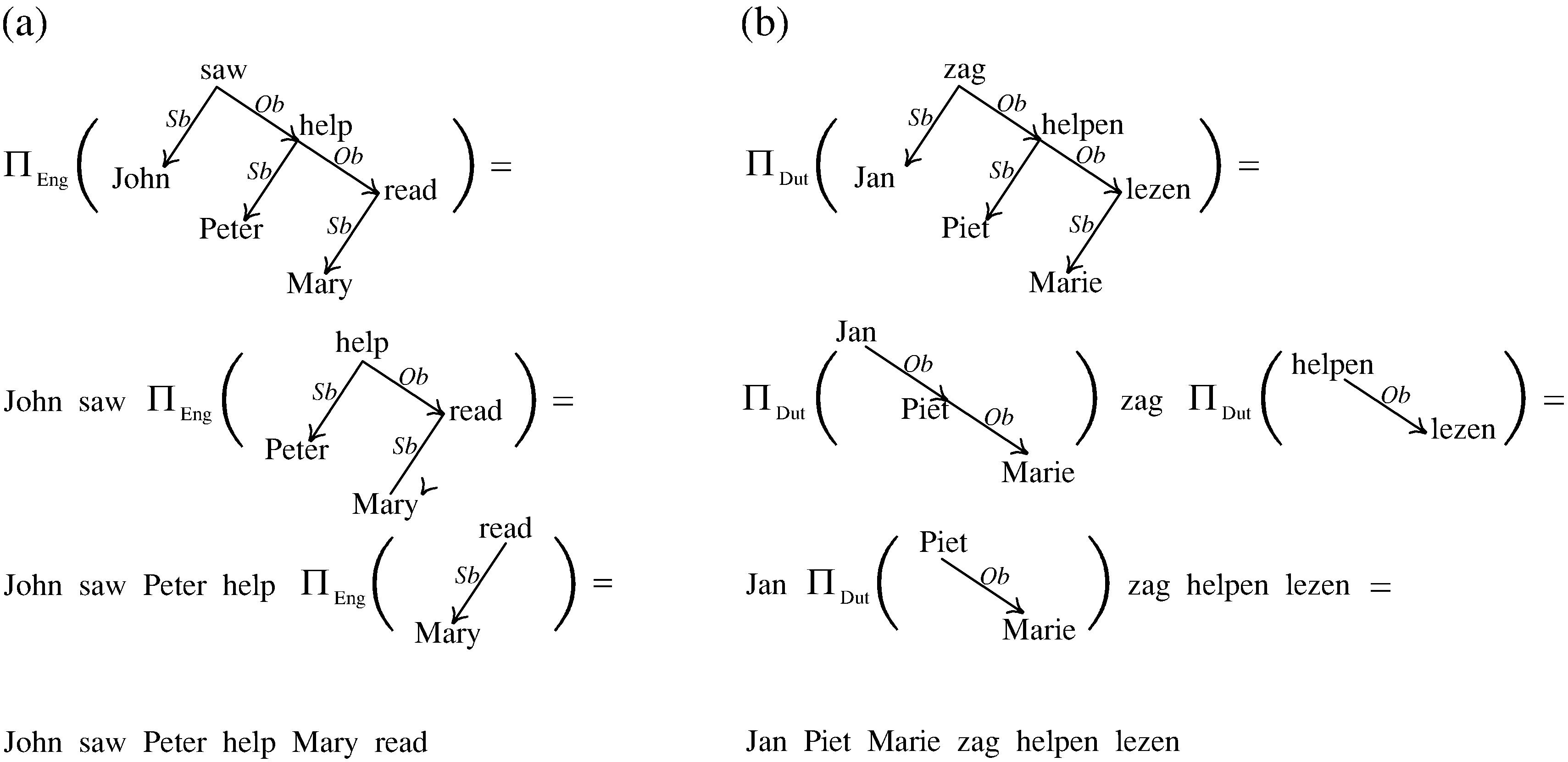}
\caption{(a) Linearization for subordination in English. (b) Linearization for subordination in Dutch.}\label{Fig11}
\end{figure}

As an illustration, let us resume the first sentences (\ref{English}) \emph{\ldots that John saw Peter help Mary read}; and the Dutch version (\ref{Dutch}) \emph{\ldots \ dat Jan Piet Marie zag helpen lezen}, which exhibits cross-serial dependencies. We assume that both sentences must have the same tree shape. Therefore the linearisations must be different for each language. Indeed, consider the two linearisations:
\begin{align}
\Pi_{\mathrm{Eng}}(S) &=\Pi_{\mathrm{Eng}}(S|\Sb) \cdot \Root(S) \cdot\Pi_{\mathrm{Eng}}(S|\Ob),\\
\Pi_{\mathrm{Dut}}(S) &=\Pi_{\mathrm{Dut}}(\Sb|S) \cdot \Root(S) \cdot\Pi_{\mathrm{Dut}}(\Ob|S).
\end{align}
Figure~\ref{Fig11}(a) and Figure~\ref{Fig11}(b) show the derivations. The first is a projective linearisation, while the second is anti-projective. We only need to swap the tree operators to obtain one from the other. We have, indeed, the relations:
$\Pi_{\mathrm{Dut}}= \overline{\Pi}_{\mathrm{Eng}}$,
or equivalently,
$\Pi_{\mathrm{Eng}}= \overline{\Pi}_{\mathrm{Dut}}$.
All this indicates that word-ordering in both languages should be considered at the same level of complexity.
In the same way, recall sentences (\ref{frasescoordinacio})a, (\ref{frasescoordinacio})b, (\ref{frasescoordinacio2})a, and (\ref{frasescoordinacio2})b, we can construct a linearisation for coordination in the standard  order, say $\Pi_{\mathrm{sta}}$, which is projective, and other linearisation, $\Pi_{\mathrm{resp}}$, anti-projective, for respectively construction. The same relation of duality holds $\Pi_{\mathrm{sta}}=\overline{\Pi}_{\mathrm{resp}}$.\footnote{ \cite{cardo2016algebraic} considered the possibility of changing the syntactic structure but not the linearisation, that is, taking locally defined dependency trees for English and anti-locally for Dutch, and a projective linearisation for both. According to Theorem~\ref{Teorema2}, there is an algebraic equivalence.}

The class of anti-context-free languages opens some questions.
It can be proved that $n$-copy languages, $\Lncopy=\{ x^n \,|\, x \in \Sigma^*\}$, are anti-context-free, and it is known that TAG only can generate the copy language $\Ltcopy$.
We do not know whether some well-known formalisms can generate the class $-\CF$. We notice that dual languages are equal up to reordering the letters of each string. That makes us suspect that the mix language could be not in $-\CF$.

It is worth considering a straightforward extension of the class $\CF \cup -\CF$ with more linguistic interest. Let us denote $\pm \PR$ the class of linearisations that can combine subtrees and anti-subtrees. Then, $\CF \cup -\CF \subseteq \LC/ \pm \PR$. However, this class could still fall short of natural languages.\footnote{There are constructions whose dependency trees are not local.
Consider the English sentences: \emph{She is an actress}, \emph{She wants to be an actress}, \emph{She wants to try to be an actress}, and so on. Since the subject \emph{she} and the predicate \emph{actress} must agree on gender, and one can put more and more material between them, the agreement is not local in the tree.}

Another question is whether ‘anti’ operators (anti-subtree, anti-class, \ldots) can be generalized to more general dependency trees, i.e., trees where a node can govern two or more children with the same syntactic function. Further work should tackle this methodological aspect jointly with the issue of coordination.

A final idea, a little more philosophical, is suggested by the very notion of duality. The fact that two objects are dual entails that neither is privileged over the other. From an algebraic perspective, this is the case for the classes $\CF$ and $-\CF$. Why has human language favoured one of the forms? To discover the causes of this symmetry breaking (frequent in other physical and biological systems), we should evaluate other linguistic aspects beyond the algebraic point of view, such as the efficiency of parsing or evolutionary syntax.
In this respect, some literature \cite{temperley2018minimizing, gomez2019memory, gomez2017scarcity} considers, as a universal principle, that natural languages have evolved to minimise the length of dependencies in a sentence. If two vertices form a dependency, its \emph{dependency length} is the distance of the vertices in the linear order, and the \emph{total length} is the sum of all dependency lengths. The sentence (\ref{English}) has total length 7, while the sentence (\ref{Dutch}), 11.
We can formalise a bit more that question in our own terms.
Fix a tree language $W$ and a linearisation $\Pi$. Given $x=\Pi(S)$, let $\ell(x)$ denote the total dependency length of the string $x$ with dependencies given by $S \in W$.
It is easily seen that $\ell$ grows linearly for $L_{\mathrm{squa}}=\Pi_{\mathrm{squa}}(W_{\mathrm{squa}})$:
\begin{align*}
\ell(x)=\frac{3|x|}{2}-2,
\end{align*}
and that it grows quadratically for $L_{\mathrm{copy}}=\overline{\Pi}_{\mathrm{squa}}(W_{\mathrm{squa}})$:
\begin{align*}
\ell(x)=\frac{|x|^2}{4}+\frac{|x|}{2}-1.
\end{align*}
See Figures~\ref{FigDependencyLength} (a) and (b). A direction of further research is to study the dependency length growth concerning the classes of tree languages and linearisations viewed in this article.

\begin{figure} 
 \begin{center}
{\small (a)} \begin{dependency}[arc edge, arc angle=60, text only label, label style={above}]
\begin{deptext}[column sep=1.2cm]
 $a$  \& $a$ \&$b$  \&   $b$  \&  $c$   \&  $c$  \\
\end{deptext}
\depedge{2}{1}{1}
\depedge{2}{4}{2}
\depedge{4}{3}{1 \quad}
\depedge{4}{6}{2}
\depedge{6}{5}{1 \quad}
\end{dependency}
\end{center}

 \begin{center}
{\small (b)} \begin{dependency}[arc edge, arc angle=60, text only label, label style={above}]
\begin{deptext}[column sep=1.2cm]
 $a$  \& $b$ \&$c$  \&   $a$  \&  $b$   \&  $c$  \\
\end{deptext}
\depedge{4}{5}{1 \quad\quad}
\depedge{5}{6}{1 \quad\quad}
\depedge{4}{1}{3}
\depedge{5}{2}{3}
\depedge{6}{3}{3}
\end{dependency}
\end{center}

\caption{(a) A string in $L_{\mathrm{squa}}=\Pi_{\mathrm{squa}}(W_{\mathrm{squa}})$ with total dependency length 7 (b) A string in $L_{\mathrm{copy}}=\overline{\Pi}_{\mathrm{squa}}(W_{\mathrm{squa}})$ with total dependency length 11.}\label{FigDependencyLength}
\end{figure}



%
%
%

\bibliographystyle{amsplain}

\begin{thebibliography}{99}


\bibitem{aho1968indexed}
  Aho, A. V.:
  {Indexed grammars, an extension of context-free grammars},
  {Journal of the ACM (JACM)},
  \textbf{15},
  {4},
  {647--671},
  {1968}.
  
  
\bibitem{autebert1997context}
  Autebert, J.M., Berstel, J. and Boasson, L.:
  {Context-free languages and pushdown automata},
  {Handbook of formal languages: Word, Language, Grammar},
  \textbf{1},
  {111--174},
  Springer, {1997}.



\bibitem{bach1981discontinous}
  Bach, E.:
  {Discontinuous constituents in generalized categorial grammar},
  {Proceedings of the 11th Annual Meeting of the North Eastern Linguistics Society},
  {1--12},
  {1981}.

\bibitem{bach1988categorial}
  Bach, E.:
  {Categorial grammars as theories of language},
  {Studies in Linguistics and Philosophy},
  \textbf{32},
  {17--34},
  {1988}.



\bibitem{bar1960finite}
  Bar-Hillel, Y. and Shamir, E.:
  {Finite-state languages: formal representations and adequacy problems},
  {Bull. of Res. Council of Israel},
  \textbf{8F}, {155--166},
  {1960}.




\bibitem{berstel2013transductions}
  Berstel, J.:
  {Transductions and context-free languages},
  Leitf\"{a}den der angewandten Mathematik und Mechanik, Teubner Studienb\"{u}cher:    Informatik,
  \textbf{38},
  {Teubner},
  2013.



\bibitem{bresnan1987cross}
  Savitch, W. J., Bach, E., Marsh, W. and Safran-Naveh, G.:
  {Cross-serial dependencies in {D}utch},
  {The formal complexity of natural language},
  {286--319},
  {1987}.


\bibitem{blum1999greibach}
  Blum, N. and Koch, R.:
  {Greibach normal form transformation revisited},
  {Information and Computation},
  \textbf{150},
  {1},
  {112--118},
  {1999}.




\bibitem{cardo2016algebraic}
  Card{\'o}, C.:
  {Algebraic Governance and Symmetry in Dependency Grammars},
  {20th and 21st International Conferences on Formal Grammar, (FG'15), revised Selected Papers, Barcelona, Catalunya},
  {60--76},
  {2016}.


\bibitem{cardo2018algebraic}
  Card{\'o}, C.:
  {Algebraic dependency grammar},
  {Universitat Polit{\`e}cnica de Catalunya}
  2018

\bibitem{pmlr-v21-clark12a}
  Clark, A. and Yoshinaka, R.:
  {Beyond Semilinearity: Distributional Learning of Parallel Multiple Context-free Grammars},
  {Proceedings of the 11th International Conference on Grammatical Inference (ICGI'12), Washington, D.C.,  USA},
  {84--96},
  {2012}.



\bibitem{comon2007tree}
   Comon, H., Dauchet, M.,  Gilleron, R., L\"oding, C., 
Jacquemard, F., Lugiez, D., Tison, D., and Tommasi, M.:
  {Tree automata techniques and applications},
   year={2007}, \url{http://tata.gforge.inria.fr/}
           [Accessed: 05/6/2020].

\bibitem{chomsky2002syntactic}
  Chomsky, N.: 
  \textit{Syntactic structures},
  Janua linguarum, Series minor, 4, Mouton \& Co., 1957. 

\bibitem{chomsky1963algebraic}
  {Chomsky, N. and Sch{\"u}tzenberger, M. P.}:
  {The algebraic theory of context-free languages},
  {Studies in Logic and the Foundations of Mathematics},
  \textbf{35},
  {118--161},
  {1963}.


\bibitem{gaifman1965dependency}
  Gaifman, H.:
  {Dependency systems and phrase-structure systems},
  {Information and control},
  \textbf{8},
  {3},
  {304--337},
  {1965}.

\bibitem{gazdar1988applicability}
  Gazdar, G.:
  {Applicability of indexed grammars to natural languages}.
  \textit{Natural language parsing and linguistic theories},
  {Studies in Linguistics and Philosophy},
  \textbf{35},
  {69--94},
  Springer, {1988}.
 
 


\bibitem{gomez2017scarcity}
  G{\'o}mez-Rodr{\'\i}guez, C. and Ferrer-i-Cancho, R.:
  {Scarcity of crossing dependencies: A direct outcome of a specific constraint?},
  {Physical Review E},
  \textbf{96},
  {6},
  {2017}.

  
\bibitem{gomez2019memory}
  G{\'o}mez-Rodr{\'\i}guez, C. , Morten H. and Ferrer-i-Cancho, R.:
  {Memory limitations are hidden in grammar},
  {arXiv preprint arXiv:1908.06629},
  {2019}.
  
  
\bibitem{gross1964equivalence}
  Gross, M.:
  {On the equivalence of models of language used in the fields of mechanical translation and information retrieval},
 {Information storage and retrieval},
  \textbf{2},
  {1},
  {43--57},
  {1964}.




\bibitem{hays1964dependency}
  Hays, D. G.: 
  {Dependency theory: A formalism and some observations},
  {Language},
  {511--525},
  {1964}.

\bibitem{hopcroft1979ANTICintroduction}
  Hopcroft, J. and Ullman, J. D.:
  {Introduction to automata theory, languages, and computation},
   Addison Wesley,{1979}.

\bibitem{hopcroft1979introduction}
  Hopcroft, J. E., Motwani R., and Ullman, J. D.:
  {Introduction to automata theory, languages, and computation},
  {Addison Wesley},
  {Pearson education},
  {2nd}, 2001.



\bibitem{joshi1985tree}
  Joshi, A. K.:
  {Tree adjoining grammars: How much context-sensitivity is required to provide reasonable structural descriptions?},
  Natural Language Parsing, 206--250, 1985.







\bibitem{kac1987simultaneous}
  Kac, M. B., Manaster-Ramer, A., and Rounds, W. C.:
  {Simultaneous-distributive coordination and context-freeness},
  {Computational Linguistics},
  \textbf{13},
  {1-2},
  {25--30},
  {1987}.


\bibitem{kallmeyer2010parsing}
  Kallmeyer, L.:
  \textit{Parsing beyond context-free grammars},
  publisher={Springer},
  {Berlin, Heilderberg,2010}


\bibitem{kanazawa2012mix}
  Kanazawa, M. and Salvati, S.:
  {{MIX} is not a tree-adjoining language},
  Proceedings of the 50th Annual Meeting of the Association for Computational Linguistics (ACL'12), Jeju, Korea.
  \textbf{1},{666--674},
  {2012}.
  
  
\bibitem{knuutila1993inference}
  Knuutila, T.:
  {Inference of k-testable tree languages},
  {Advances in Structural and Syntactic Pattern Recognition: proceedings of the international workshop (SSPR'92), Bern, Switzerland},
  {Machine perception artificial intelligence},
  \textbf{5},
  {109--120},
  {1993}.
 
\bibitem{kuhlmann2010dependency}
  Kuhlmann, M.:
  {Dependency Structures and Lexicalized Grammars: An Algebraic Approach},
  \textbf{6270},
  {Published PhD thesis},
  {Springer}, 2010.
  
  
\bibitem{melcuk1988dependency}
  Mel{\v{c}}uk, I. A.:
  {Dependency Syntax: Theory and practice},
  sunny series in linguistics, State University of New York Press, 1998.
  

\bibitem{nivre2005dependency}
  Nivre, J.:
  {Dependency grammar and dependency parsing},
  {V\"{a}xj\"{o} University: School of Mathematics and Systems Engineering},
  {MSI 05133},
  {5133},
  {1--32},
  {2005}.



\bibitem{parikh1966context}
  Parikh, R. J.:
  {On context-free languages},
  {Journal of the ACM (JACM)},
  \textbf{13},
  {4},
  {570--581},
  {1966}.

\bibitem{pullum1982natural}
  Pullum, G. K. and Gazdar, G.:
  Natural languages and context-free languages.
  Linguistics and Philosophy, \textbf{4}, 471--504, 1982.



\bibitem{popel2013coordination}
  Popel, M., Marecek, D., Step{\'a}nek, J., Zeman, D. and Zabokrtsk{\`y}, Z.:
  {Coordination Structures in Dependency Treebanks},
  {Proceedings of the 13rd conferences of the association for computational linguistics (ACL'13), Sofia, Bulgaria},
  pages={517--527},
  year={2013}.



\bibitem{robinson1970dependency}
  Robinson, J. J.:
  {Dependency structures and transformational rules},
  {Language},
  pages={259--285},
  \textbf{46},
  {1970}.





\bibitem{seki1991multiple}
  Seki, H., Matsumura, T., Fujii, M. and Kasami, T.:
  {On multiple context-free grammars},
  {Theoretical Computer Science},
  \textbf{88},
  {2},
  {191--229},
  {1991}.


\bibitem{shieber1987evidence}
  Shieber, S. M.:
  Evidence against the context-freeness of natural language.
  Linguistics and philosophy,
  \textbf{8}, {333--343}, {1985}.



\bibitem{temperley2018minimizing}
  Temperley, D. and Gildea, D.:
  {Minimizing syntactic dependency lengths: Typological/cognitive universal},
  {Annual Review of Linguistics},
  \textbf{4},
  {1},
  {67--80},
  {2018}.

\bibitem{tesniere1959elements}
  Tesni{\`e}re, L.:
  {El{\'e}ments de syntaxe structurale},
  {Klincksieck}, 
  {Paris}, {1959}.



\bibitem{vijay1987characterizing}
  Vijay-Shanker, K., Weir, D. J. and Joshi, A. K.:
  {Characterizing structural descriptions produced by various grammatical formalisms},
  Proceedings of the 25th annual meeting on Association for Computational Linguistics,
  {104--111},
  {1987}.





\bibitem{weir1988combinatory}
  Weir, D. and Joshi, A.:
  {Combinatory categorial grammars: Generative power and relationship to linear context-free rewriting systems},
  {26th Annual Meeting of the Association for Computational Linguistics},
  {278--285},
  {1988}.





\bibitem{yokomori1995polynomial}
  Yokomori, T.:
  {On polynomial-time learnability in the limit of strictly deterministic automata},
  {Machine Learning},
  \textbf{19},
  {2},
  {153--179},
  {1995}.


\bibitem{zalcstein1972locally}
  Zalcstein, Y.:
  {Locally testable languages},
  {Journal of Computer and System Sciences},
  \textbf{6},
  {2},
  {151--167},
  {1972}.

\end{thebibliography}

\end{document}